\documentclass[twocolumn,trackchanges]{aastex62}
\usepackage{hyperref}
\hypersetup{bookmarksnumbered=true, bookmarksopen=true}

\usepackage{amssymb,amsmath,amsfonts}
\usepackage{footnotebackref}

\allowdisplaybreaks[1]

\usepackage{savesym}
\savesymbol{tablenum}
\usepackage{siunitx}
\restoresymbol{SIX}{tablenum}

\usepackage{bm}
\usepackage{booktabs}
\usepackage{paralist}

\usepackage{newtxtext}
\usepackage[varvw]{newtxmath}
\DeclareSymbolFont{CMletters}{OML}{cmm}{sb}{it}
\DeclareMathSymbol{\nu}{\mathord}{CMletters}{23}

\graphicspath{{figs3/}}

\newcommand{\refsec}[1]{Section~\ref{#1}}
\newcommand{\reffig}[1]{Figure~\ref{#1}}
\newcommand{\reftab}[1]{Table~\ref{#1}}
\newcommand{\refeqn}[1]{Equation~(\ref{#1})}
\newcommand{\refeqnalt}[1]{Equation~\ref{#1}}

\newcommand{\dif}{\mathop{}\!\mathrm{d}}
\newcommand{\me}{\mathrm{e}}

\newcommand{\abs}[1]{\left\lvert #1 \right\rvert}    
\newcommand{\avg}[1]{\left\langle #1 \right\rangle}  

\renewcommand{\vr}{v_{\mathrm{r}}}
\newcommand{\vt}{v_{\mathrm{t}}}

\newcommand{\mh}{M_{\mathrm{h}}}
\newcommand{\rh}{R_{\mathrm{h}}}
\newcommand{\vh}{V_{\mathrm{h}}}

\newcommand{\rhoc}{\rho_{\mathrm{crit}}}

\newcommand{\mpc}{\mathrm{Mpc}}

\newcommand{\kms}{\mathrm{km \, s}^{-1}}
\newcommand{\msunh}{h^{-1} M_{\odot}}
\newcommand{\mpch}{h^{-1} \mathrm{Mpc}}
\newcommand{\kpch}{h^{-1} \mathrm{kpc}}

\newcommand{\LCDM}{\ensuremath{\Lambda\mathrm{CDM}}}

\newcommand{\cossq}{\cos^2\theta}

\shorttitle{ORBITAL DISTRIBUTION OF INFALLING SUBHALOS}
\shortauthors{Li et al.}

\begin{document}

\title{\large\textbf{Orbital distribution of infalling satellite halos across cosmic time}}


\author[0000-0001-7890-4964]{Zhao-Zhou Li}
\affil{Department of Astronomy, School of Physics and Astronomy, Shanghai Jiao Tong University,
  Shanghai 200240, China; \href{mailto:lizz.astro@gmail.com}{lizz.astro@gmail.com}}
\affil{Key Laboratory for Research in Galaxies and Cosmology, Shanghai Astronomical Observatory, 
  Shanghai 200030, China; \href{mailto:dhzhao@shao.ac.cn}{dhzhao@shao.ac.cn}}
\affil{Shanghai Key Laboratory for Particle Physics and Cosmology, Shanghai 200240, China}

\author{Dong-Hai Zhao}
\affil{Key Laboratory for Research in Galaxies and Cosmology, Shanghai Astronomical Observatory, 
  Shanghai 200030, China; \href{mailto:dhzhao@shao.ac.cn}{dhzhao@shao.ac.cn}}

\author[0000-0002-4534-3125]{Y. P. Jing}
\affil{Department of Astronomy, School of Physics and Astronomy, Shanghai Jiao Tong University,
  Shanghai 200240, China; \href{mailto:lizz.astro@gmail.com}{lizz.astro@gmail.com}}
\affil{Tsung-Dao Lee Institute, Shanghai Jiao Tong University, Shanghai 200240, China}
\affil{Shanghai Key Laboratory for Particle Physics and Cosmology, Shanghai 200240, China}

\author[0000-0002-8010-6715]{Jiaxin Han}
\affil{Department of Astronomy, School of Physics and Astronomy, Shanghai Jiao Tong University,
  Shanghai 200240, China; \href{mailto:lizz.astro@gmail.com}{lizz.astro@gmail.com}}
\affil{Shanghai Key Laboratory for Particle Physics and Cosmology, Shanghai 200240, China}

\author{Fu-Yu Dong}
\affil{Key Laboratory for Research in Galaxies and Cosmology, Shanghai Astronomical Observatory, 
  Shanghai 200030, China; \href{mailto:dhzhao@shao.ac.cn}{dhzhao@shao.ac.cn}}
\affil{School of Physics, Korea Institute for Advanced Study (KIAS), 85 Hoegiro, Dongdaemun-gu, Seoul, 02455, Korea}

\begin{abstract}

The initial orbits of infalling subhalos largely determine 
the subsequent evolution of the subhalos and satellite galaxies therein
and shed light on the assembly of their hosts.
Using a large set of cosmological simulations of various resolutions,
we quantify the orbital distribution of subhalos at infall time
and its mass and redshift dependence in a large dynamic range.
We further provide a unified and accurate model validated across cosmic time,
which can serve as the initial condition for semi-analytic models.
We find that the infall velocity $v$ follows a nearly universal distribution peaked near the host virial velocity $V_{\mathrm{h}}$
for any subhalo mass or redshift,
while the infall orbit is most radially biased when $v\sim V_{\mathrm{h}}$.
Moreover, subhalos that have a higher host mass or a higher sub-to-host ratio 
tend to move along a more radial direction 
with a relatively smaller angular momentum than their low host mass or low sub-to-host ratio counterparts, 
though they share the same normalized orbital energy.
These relations are nearly independent of the redshift 
when using the density peak height as the proxy for host halo mass. 
The above trends are consistent with the scenario where the dynamical environment is relatively colder for more massive structures
because their own gravity is more likely to dominate the local potentials.
Based on this understanding, the more massive or isolated halos are expected to have higher velocity anisotropy.

\end{abstract}


\section{Introduction} \label{sec:intro}

In the current hierarchical structure formation framework, 
dark matter halos grow through the accretion of diffuse matter and smaller halos.
After infall, the subhalos will experience dynamical friction,
tidal heating and stripping,
and ultimately get dissolved
into the hosts 
(see, e.g., \citealt{Zavala2019} and references therein).
Meanwhile, the inhabiting satellite galaxies might undergo
morphology transformation arising from ram-pressure stripping, strangulation, harassment,
and tidal shock heating;
the interactions and mergers, especially major mergers, can drive the evolution of the central galaxy (\citealt{Mo2010}).
The efficiency and the outcome of all of these processes depend critically on the details of subhalo orbits.
On the other hand, the subhalos are building blocks of larger halos;
their orbital distribution 
can help us to understand the phase-space structure of halos and the environment dependence,
including the mass profile \citep{Dalal2010}, angular momentum \citep{Vitvitska2002,Bett2012,Benson2020}, 
and velocity anisotropy \citep{Ludlow2011,Sparre2012,Shi2015}.
Moreover, the satellite kinematics are widely used to infer the potential of their hosts \citep[e.g.,][]{Diaferio1997,More2011,Li2017,Li2019a}
or model the galaxy clustering in redshift space \citep[e.g.,][]{Blake2011,Zu2013,Shi2016}.
Therefore, the cosmological predictions for the orbital distribution of subhalos
are indispensable to modeling galaxy formation, understanding halo structure, and interpreting observation.

The orbital distribution of subhalos at the time first entering their host halo
are of particular interest. 
It represents the initial condition, which largely determines the subsequent evolution.
Hence, it becomes an important ingredient of semi-analytical models (e.g., \citealt{Baugh2006,Yang2011,Jiang2020}).
In addition, from a practical view,
the pre-merger subhalos in numerical simulations are more robust against 
the discrepancy among subhalo finding algorithms (\citealt{Han2012,Han2018a,Onions2012,Behroozi2013b}),
the artificial disruption due to insufficient resolution \citep{Han2016,vandenBosch2018a},
or the baryonic process (e.g., \citealt{Zhu2016a,Sawala2017,Richings2018}).
Because these subhalos retain higher mass, reside in less dense environments,
and have barely experienced the complicated dynamical interactions inside hosts
compared to their evolved descendants.
For this reason, semi-analytical models based on these initial conditions
can outperform simulations in terms of numerical resolution,
and thus become valuable complementary tools for its flexibility \citep{Jiang2020}.

In the idealized scenario of spherical collapse (\citealt{Gunn1972}, see also \citealt{Mo2010}), 
assuming purely radial orbits and the conservation of energy,
a mass shell first expands until reaching the turnaround radius $R_\mathrm{ta}$,
then falls back onto the virialized halo of size $\rh \simeq 0.5 R_\mathrm{ta}$ and mass $\mh$.
Given
$- \frac{G \mh}{R_\mathrm{ta}} = - \frac{G \mh}{\rh} + \frac{1}{2}v^2$,
the infalling materials (including subhalos) are expected to move radially 
with velocity $v \simeq \sqrt{G\mh/\rh} \equiv \vh$
when they first cross the host virial radius $\rh$.
This is clearly oversimplified, 
because the overdensity regions are generally aspherical and perturbed by surrounding structures,
which can introduce random motions.

More realistic descriptions have been studied using cosmological simulations.
People find that subhalos at infall time indeed move with velocities around $\vh$
but have intermediate orbital circularities, 
i.e., very radial or tangential orbits are both relatively rare \citep[e.g.,][]{Tormen1997,Vitvitska2002}.
The subhalos tend to follow more radial orbits with smaller scaled angular momentum and pericenter distance
for more massive hosts \citep{Wetzel2011,Jiang2015}
or higher sub-to-host mass ratio \citep{Tormen1997,Wang2005,Jiang2015}.
Their position vectors are aligned with the shape of the hosts
and with the principal axes of external tidal fields \citep[e.g.,][]{Benson2005,Wang2005,Libeskind2014,Shi2015,Kang2015a,Wang2017a},
in particular, the tangential velocity increases significantly with the strength of the tidal field \citep{Shi2015}.

An accurate depiction of the mass and redshift dependence for subhalo orbits
is essential to model the halo assembly and galaxy formation across cosmic time.
However, most works have focused on the host halos at $z=0$ (e.g., \citealt{Wang2005,Jiang2015})
or at very few time snapshots \citep{Benson2005},
and the limited sample size and dynamic ranges have inhibited precisely characterizing the mass and redshift dependence.
\citet{Wetzel2011} reported that the subhalo orbits become more radial at higher redshift for given halo mass,
and provided fitting formulae to the orbital circularity and pericenter distance as a function of the host mass and redshift.
Nevertheless, as pointed out by  \citet{Jiang2015},
\citeauthor{Wetzel2011} was unable to detect the dependence on the sub-to-host ratio
and did not examine the correlation between the two orbital parameters.

In this paper, we aim to provide a comprehensive and unified description of the orbits of infalling subhalos
\textit{across cosmic time}.
Using merging halo pairs from 16 cosmological simulations of various resolutions,
the unprecedented large sample size and dynamic range allow us to
characterize the joint distribution of orbital parameters and 
the mass and redshift dependence with high precision.
While the general trend is qualitatively consistent with previous discoveries,
we have much better statistics for the cases
that are important but generally poorly covered in previous studies,
including massive cluster halos, major mergers, or halos at high redshift.
Moreover, we find a clear dependence of the infall angle on the velocity for the first time.
The infall angle is most radially biased when $v\sim\vh$,
which is contrary to the assumption made by \cite{Jiang2015}.

It would be convenient to have analytical formulae for orbital distribution.
However, the large parameter space makes any unified description intractable,
especially considering the nontrivial redshift dependence reported by \citet{Wetzel2011}.
Inspired by the halo formation theory (e.g., \citealt{Bardeen1986,Sheth2001,Mo2010}),
we find that using the peak height $\nu$ as a proxy for host mass can remove the redshift (and likely the cosmology) dependence,
which hence allows us to build an accurate but simple model.

The structure of this paper is as follows.
We first introduce the simulations and merging halos in \refsec{sec:data}.
We then present our model for the orbital distribution in \refsec{sec:model},
and verify the model with simulation data in \refsec{sec:result}.
We discuss the results in \refsec{sec:discussion} and summarize in \refsec{sec:conclusion}.

We adopt the host halo's center and velocity as the reference frame for subhalo kinematics through this paper. 
The Hubble flow is included in subhalo velocity.

\section{Data and method}\label{sec:data}

\subsection{Simulations and halos} \label{sec:simu}

This work is based on a set of high-resolution $N$-body simulations \citep{Jing2007}.
It contains 16 realizations of three resolutions
that were carried out with the parallel particle--particle--particle--mesh ($\rm{P^3M}$) code \citep{Jing2002} 
under a flat $\LCDM$ cosmology with 
$\Omega_\mathrm{m} = 0.268$, $\Omega_{\Lambda} = 0.732$, $H_0= 71\, \kms \mpc^{-1}$, $n_\mathrm{s}=1$, $\sigma_8 = 0.85$.
Though the parameters are not the most up to date, 
our results are possibly not very sensitive
to the cosmological parameters, as we argue in \refsec{sec:cosmo}.
Each realization contains $1024^3$ particles within periodic boundaries
and outputs a series of snapshots equally spaced in $\log(a)$.
The detailed parameters, including the box size, resolution, and output redshifts, are listed in \reftab{tab:simu}.
The realizations belong to three different resolutions that are labeled as L150, L300, and L600, respectively,
according to the box size in units of $\mpch$.
Such a configuration not only enlarges the sample size and dynamic range greatly
but also enables better discrimination of the numerical effects.

\begin{table*}[htb]
  \caption{Simulation parameters.
  The simulations are labeled into three groups by resolution.
  The columns present the label of the group, number of realizations, box size, number of particles, particle mass, softening length, 
  the redshift range, and number of output snapshots, respectively.
  }
  \label{tab:simu}
  \centering
  \setlength{\tabcolsep}{12pt}
  \begin{tabular}{cccccccc}
  \toprule
  \toprule
    Label & $N_\mathrm{realization}$ & $L_\mathrm{box}$ & $N_\mathrm{p}$  & $m_\mathrm{p}$  & $\epsilon_\mathrm{soften}$ & $z_\mathrm{snap}$ & $N_\mathrm{snap}$\\
          &        & $\mpch$          &                 & $\msunh$        & $\kpch$                    &                   &\\ 
  \midrule
    L150  & 8 & 150 & $1024^3$ & \num{2.34e8}  & 5  & $[0.0, 16.9]$ & 100\\
    L300  & 4 & 300 & $1024^3$ & \num{1.87e9}  & 10 & $[0.0, 16.9]$ & 100\\
    L600  & 4 & 600 & $1024^3$ & \num{1.50e10} & 30 & $[0.0, \phantom{0}7.3]$  & \phantom{0}24\\
  \bottomrule
  \end{tabular}
  \vspace{1em}
\end{table*}

For each simulation snapshot, 
we identify the halos using the standard Friends-of-Friends (FoF, \citealt{Davis1985}) algorithm
with a linking length equal to $0.2$ times the mean particle separation.
The center and velocity of a halo are specified by its largest subhalo (see below).
The virial mass $\mh$ and radius $\rh$ are defined as the quantities of a spherical region enclosing the halo center
with a mean density equal to $\Delta_\text{vir}\rhoc$,
where $\Delta_\text{vir}$ is the virial factor \citep{Bryan1998},
and $\rhoc$ is the critical density of the universe at the snapshot.
We characterize the halo mass across cosmic time
by the peak height, $\nu(\mh, z)$, 
which is the natural unit in the \citet{Press1974} formalism (see also \citealt{Bond1991}, \citealt{Mo2010}).
The peak height is defined as $\nu=\delta_\text{col}(z)/\sigma(\mh,z)$,
where $\delta_\text{col}(z)=1.686\,\Omega_m(z)^{0.0055}$ is the critical overdensity required for spherical collapse at $z$
and $\sigma^2(\mh,z)$ is the variance of the density field smoothed on scale of $\mh$ predicted by linear theory.\footnote{
  Interested readers can use the open-source code Colossus \citep{Diemer2017a}
  at \url{https://bitbucket.org/bdiemer/colossus/} to calculate $\nu$ as a function of halo mass, redshift, and cosmological parameters.
}
At a given redshift, 
$\nu$ is a monotonic function of the halo mass (see \reffig{fig:n_nu_xi}).

We apply the Hierarchical Bound-Tracing (HBT, \citealt{Han2012})
algorithm to search subhalos and build merger trees.\footnote{
  An updated version, HBT$+$ \citep{Han2018a},
  is publicly available online at \url{https://github.com/Kambrian/HBTplus/}.
  It has improved the speed, the user interface, and the physical treatment of trapped subhalos
  that have sunk to the center of their hosts but not been disrupted.
}
HBT is considered to be the \textit{state of the art} code for finding and tracing subhalos
according to various systematical tests \citep{Han2012,Han2018a,Onions2012,Srisawat2013}.
Unlike the conventional algorithms like SUBFIND \citep{Springel2001} that search subhalos at individual snapshots separately,
HBT identifies (sub)halos as they form, tracks their evolution as they merge, and builds their merger trees simultaneously,
hence performing reliably even in the very dense background of a host halo.
The mass $m$ of a subhalo is defined as its self-bound mass,
while the position and velocity are defined 
as the average position and velocity of the most bound 25\% core particles.
This definition is more physical and robust than that based on the center of mass of all member particles (see \citealt{Han2012} for details).

To alleviate the numerical effects, we only use the subhalos with more than 60 bound particles
and halos with more than 600 particles within $\rh$.
This halo mass limit corresponds to
$\mh=\num{1.4e11}\allowbreak \msunh$ for L150 boxes ($\nu=0.6$ at $z=0$, $\nu=3.2$ at $z=6$)
and $\num{9e12} \msunh$ for L600 boxes.
We have confirmed that the results from simulations of different resolutions are consistent
when using the above selection criteria.

\subsection{Merging halo pairs} \label{sec:sample}

We consider the subhalos that \textit{for the first time} enter their host halo,
or say the halo pairs that just start to merge.
More specifically, at each snapshot, we pick all merging halo pairs that satisfy the following criteria:
\begin{compactitem}
    \item The center of subhalo is located in the virial radius of the host halo at the current snapshot,
    \item But has not been enclosed by the virial radius of \textit{this} halo at any earlier snapshot.
\end{compactitem}
Searching through all snapshots of 16 simulations, 
we find $6.6 \times 10^6$ such halo pairs in total, making a sample far larger than that used in previous related work.

We further use cubic splines to interpolate the subhalo orbits between adjoint snapshots to find the precise crossing time.
Assuming that the halo size is evolving exponentially with time,
we solve the exact redshift $z$ when the center of the subhalo reaches the virial radius
and the position and velocity of subhalo, and masses of both halos at this particular moment.
The details are given in the Appendix.
Note that a subhalo with a higher velocity tends to be found at a position deeper in the halo,
and thus shows a larger deviation from the expected distribution at infall time.
The interpolation can help to reduce such artificial effects.
Similar treatment has been widely considered in earlier works (e.g., \citealt{Benson2005,Jiang2015,Fillingham2015}).

\begin{figure}[!tbph]
  \centering
  \includegraphics[width=0.48\textwidth]{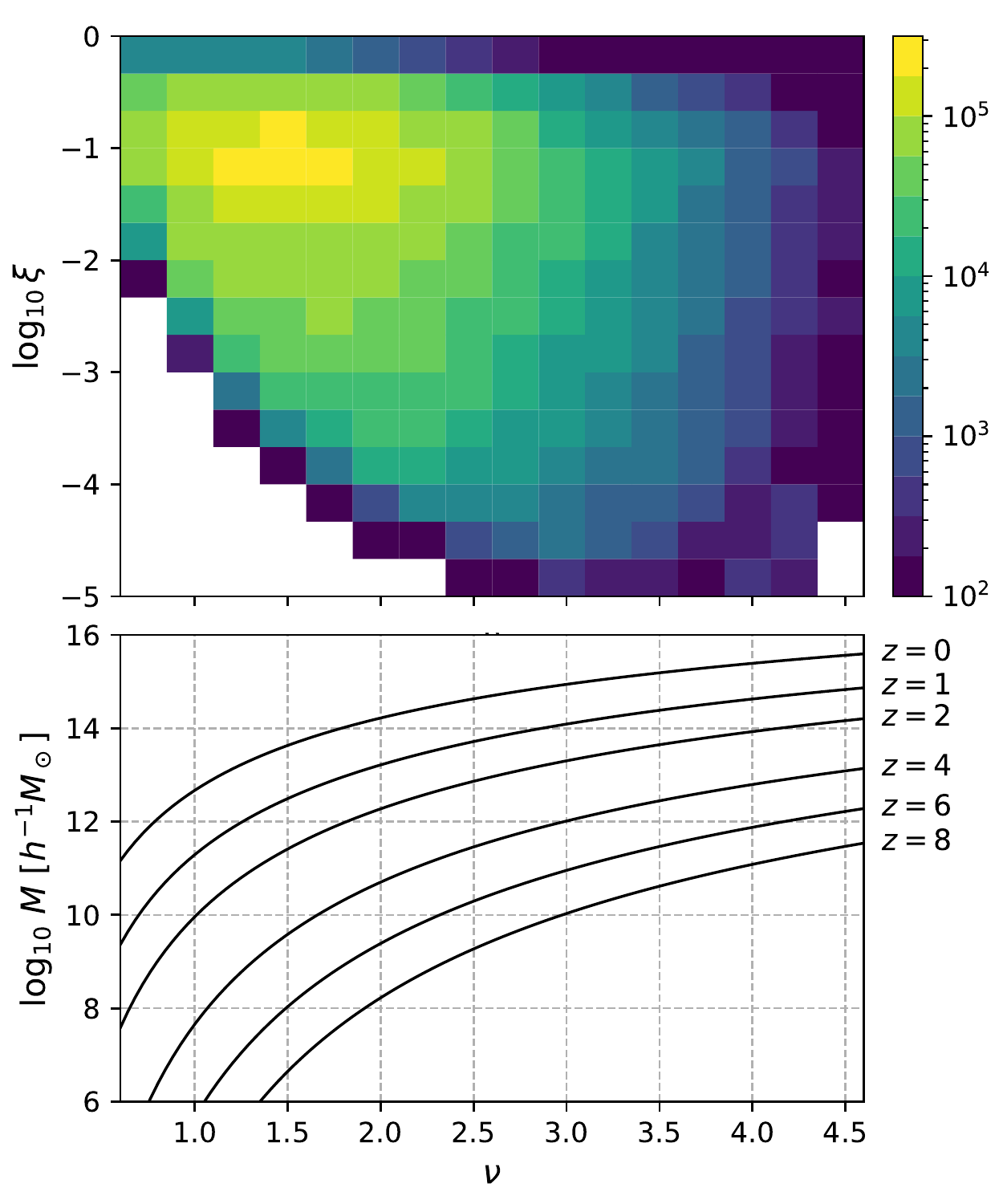}
  \vspace{-2em}
  \caption{Number of halo pairs
    binned by the host halo's peak height, $\nu$, and the sub-to-host mass ratio, $\xi=m/\mh$.
    The relations between mass and $\nu$ at various $z$ are shown in the lower panel for reference.
  }
  \label{fig:n_nu_xi}
\end{figure}

We then characterize the masses of each merging halo pair by the dimensionless quantities,
the host halo's peak height, $\nu$, and the sub-to-host mass ratio, $\xi=m/\mh$.
As demonstrated later in \refsec{sec:z}, the subhalo orbital distribution almost does not depend on the redshift 
when $\nu$ and $\xi$ are controlled.
Therefore, we can stack the halos at different epochs to further enlarge the sample
size and dynamic range for better determining the orbital distribution and possible mass dependence.
\reffig{fig:n_nu_xi} 
shows the stacked sample binned by $\nu$ and $\xi$.
The number of halos is a decreasing function of $\nu$ and $\xi$ in theory.
The actual distribution is truncated due to the mass limit
that depends on both resolution and redshift
(e.g., we can only resolve halo pairs with large $\nu$ and $\xi$ at high $z$),
thus appears non-monotonic in the figure.

It is worth pointing out the difference between the two ways of searching halos that are about to merge.
Here we consider the subhalos entering the host radius in a given time interval between adjoint snapshots,
as they are directly related to the halo growth (see also \citealt{Jiang2015}).
In early studies, people conventionally examine the neighboring (sub)halos within given distance interval $(\rh,\rh+\Delta r)$
around the hosts at single snapshot (e.g., \citealt{Tormen1997,Benson2005,Wang2005,Wetzel2011}).
It is largely because the limited resolution prevents people from identifying and tracing subhalos across time.
When using such scheme, one must apply a weighting to correct
for the underrepresentation of satellites with larger infall velocity or more radial orbit,
as they spend less time in this radial interval.
Moreover, caution should be exercised when treating the major mergers \citep{Benson2005}.
In contrast, taking advantage of merger trees, 
our sample naturally forms a complete census of all the accreted subhalos.

In the following, some more details about the sample are provided for interested readers.
A host halo at a given snapshot might occur in our sample multiple times if it is merging with
multiple subhalos simultaneously.
Conversely, a subhalo might enter into different halos successively, 
e.g., the sub-subhalos falling into a new host along with their original host (sub)halos.
The HBT algorithm has recorded the hierarchy of subhalos.
In this paper, we only use the direct subhalos in our sample (85\% of all infalling subhalos),
considering that the sub-subhalos
represent a different population due to the group preprocessing
(e.g., \citealt{Fujita2004,McGee2009,Wetzel2013,Vijayaraghavan2013,Bahe2019}).
Nevertheless, we find that the inclusion of the sub-subhalos
almost does not change the results of this paper.

We count all the newly accreted subhalos,
though they do not necessarily stay within $\rh$ afterward.
After the first infall, a subhalo with high energy may escape as a flyby \citep{Sales2007,Ludlow2009,Wang2009,Martin2020}
or, more likely, traverse the halo radius multiple times until eventually getting trapped
due to dynamical friction and mass growth of the host \citep{Balogh2000,Mamon2004,Gill2005}.
For this reason, we do not exclude high-speed subhalos.

\section{Model} \label{sec:model}

Through systematical analysis of the sample in \refsec{sec:data},
we find that the following model 
provides a comprehensive and accurate description of the orbital parameter distribution of infalling subhalos \textit{across cosmic time}.
The model validation with cosmological simulations and related discussion are given in \refsec{sec:result} and \ref{sec:discussion}.

For a spherical potential, 
the orbit of a satellite halo can be specified by two independent parameters.
Here we consider the infall velocity, $v$, and 
the angle, $\theta$, between the position vector and the velocity.
The radial and tangential velocity are thus $\vr = v\cos\theta$ and $\vt=v\sin\theta$ respectively.
As shown in \refsec{sec:deriv},
it is straightforward to transform from one set of orbital parameters to another. 

\textbullet~ 
The normalized velocity, $u=v/\vh$, of infalling subhalos follows a log-normal distribution,
\begin{equation}
    p(u)\dif u = \frac{1}{\sqrt{2\pi}\sigma_1} 
           \exp \left[{-\frac{\ln^2(u/\mu_1)}{2\sigma_1^2}} \right] \frac{\dif u}{u},
    \label{eq:v_distr}
\end{equation}
where $\vh=\sqrt{G \mh/\rh}$ is the virial velocity of the host halo.
It satisfies $\int_0^\infty p(u)\dif u=1$ for any positive $\mu_1$ and $\sigma_1$.
The median and the mode of $p(u)$ are $\mu_1$ and $\mu_1 \me^{-\sigma_1^2}$ respectively.
We find that the distribution of velocity is nearly independent of redshift or the masses of both halos.

\textbullet~ 
The infall angle in terms of $\cossq=\vr^2/v^2$ follows an exponential distribution
that depends on the velocity $u$, the peak height of the host $\nu$, and the sub-to-host mass ratio $\xi$,
\begin{equation}
    p(\cossq \mid u, \nu, \xi) \dif \cossq 
        = \frac{\eta}{\me^{\eta}-1} \exp \left({\eta \cossq} \right) \dif \cossq ,
    \label{eq:theta_distr}
\end{equation}
where  $\eta$ is a function of $u, \nu$ and $\xi$,
\begin{equation}
    \eta = a_0 \exp \left[ - \frac{\ln^2  (u / \mu_2)}{2 \sigma_1^2} \right] + A (u + 1) + B, \label{eq:eta}
\end{equation}
with $A = a_1 \nu + a_2 \xi^c + a_3 \nu \xi^c$ and $B = b_1 + b_2 \xi^c$.
It satisfies $\int_0^1 p(\cossq)\dif \cossq=1$ for any $\eta$.
We suggest taking $\eta=0$, i.e., $p(\cossq) = 1$, if a negative value is obtained from \refeqn{eq:eta}.
Here $\cossq$ is used for the simplicity of its functional form, the distribution of $\theta$ then writes
$p (\theta) = 2 \sin \theta \cos \theta p (\cos^2 \theta)$.

A larger positive $\eta$ implies that the subhalo orbits are more radial on average.
As demonstrated in \refsec{sec:phasespace}, 
$\eta=0$ represents a fully isotropic motion of infalling subhalos at the halo radius.

\textbullet~ 
The joint distribution of $u$ and $\theta$ then writes
\begin{equation}
    p(u, \theta \mid \nu,\xi) = p(u) p(\theta \mid u, \nu,\xi).
\end{equation}
Fitting the model with our halo sample,
we obtain the values of the constants in \refeqn{eq:v_distr} and (\ref{eq:eta}), 
$\mu_1$, $\sigma_1$, $\mu_2$, $a_0$, $a_1$, $a_2$, $a_3$, $b_0$, $b_2$, and $c$,
which are listed in \reftab{tab:orbit_param}.

\begin{table}[tbp]
  \caption{Constants of the orbital distribution model for \refeqn{eq:v_distr} and (\ref{eq:eta}).}
  \label{tab:orbit_param}
  \centering
  \setlength{\tabcolsep}{3.5pt}
  \begin{tabular}{cccccccccc}
    \toprule
    \toprule
    $\mu_1$ & $\sigma_1$ & $\mu_2$ & $a_0$ & $a_1$ & $a_2$ & $a_3$ & $b_1$ &
    $b_2$ & $c$\\
    \midrule
    1.20 & 0.20 & 1.04 & 0.89 & 0.30 & $-3.33$ & 0.56 & $-1.44$ & 9.60 & 0.43\\
    \bottomrule
  \end{tabular}
\end{table}

Remarkably, such a simple model can well describe the joint distribution
of orbital parameters and the mass dependence in a large dynamic range
across cosmic time.
This model is much simpler than those in the literature.
Most authors only provide fitting results in several
mass or redshift bins separately \citep[e.g.][]{Benson2005,Wang2005,Jiang2015}.
\citet{Wetzel2011} has employed up to 12 free parameters to fit the time evolution and host mass dependence,
while some more would be required if further including the subhalo mass dependence and the correlation between orbital properties.
The simplicity and accuracy of our model should be attributed to 
the use of the dimensionless variables  ($\nu$ and $\xi$) motivated by halo formation theory
and the appropriate separation of the physical components.
It also warrants further investigation of the mechanism behind.

Noticing that the best-fit parameters in \reftab{tab:orbit_param} satisfy
$\mu_1 \simeq \me^{4\sigma_1^2}$,
hence $p(u) \simeq (\sqrt{2\pi}\sigma_1 \me^{8\sigma_1^2})^{-1} u^3 \exp [ -\frac{\ln^2 u}{2\sigma_1^2} ]$.
It seems not a simple coincidence (see \refsec{sec:phasespace})
and possibly allows us to further simplify the model.

\subsection{Distribution of other orbital parameters} \label{sec:deriv}

There are various choices for the two parameters
to specify an orbit in spherical potential.
This paper (and, e.g., \citealt{Benson2005,Wang2005,Jiang2015}) uses the velocities at infall
for being simple and directly measurable.
Other choices in the literature include 
the energy, $E$, and the angular momentum \citep[e.g.,][]{Li2017},
the radius of circular orbit for given $E$ and the orbital circularity \citep{Tormen1997,vandenBosch2017},
and the pericenter distance (or orbit semi-major axis) and the eccentricity \citep{Benson2005,Wetzel2011}.
These have the advantage of depending only on the conserved quantities
but require modeling the halo potential.
In particular, the use of circularity is motivated by theoretical modeling 
to the dynamical friction \citep[e.g.,][]{Lacey1993,Jiang2008}.
It is straightforward to derive the distribution of any orbital parameters from our $p(v,\theta)$ model
through variable transformation and possibly marginalization.

For example, the joint distribution of $\vr,\vt$ is given by
\begin{equation}
  p (\vr, \vt) = \frac{1}{v} p (v, \theta),
  \label{eq:vr_vt}
\end{equation}
where $\abs{\partial{(\vr, \vt)}/\partial{(v, \theta)}}=v$ is used in the derivation.
See \reffig{fig:p_vr_vt} for an illustration.

The joint distribution of the specific orbital energy, $E=\Phi_0+\frac{1}{2}v^2$ 
and the angular momentum, $L=r\vt$, writes
\begin{equation}
  p (E, L) = \frac{1}{v^2 r \cos \theta} p (v, \theta) ,
\end{equation}
where $\Phi_0=\Phi(\rh)$ is the potential energy at $r=\rh$ and the Jacobian
$\abs{\partial{(E, L)}/\partial{(v, \theta)}}=v^2 r \cos \theta$ is used.
We can also derive the marginalized distribution.
It is straightforward to show that $E$ follows a log-normal distribution similar to velocity,
\begin{equation}
    p(E) = \frac{1}{\sqrt{2\pi}\sigma_E (E-\Phi_0)} 
           \exp \left\{{-\frac{\ln^2 \left[ (E-\Phi_0)/\mu_E \right]}{2\sigma_E^2}} \right\},
    \label{eq:E_distr}
\end{equation}
where $\mu_E=\frac{1}{2}\mu_1^2\vh^2$ and $\sigma_E=2\sigma_1$.
Meanwhile, $L$ follows
\begin{equation}
  p (L) = \int \delta(L-r v \sin \theta) p(v, \theta) \dif v \dif \theta 
        = \int_0^\frac{\pi}{2} \frac{p(v', \theta) d \theta}{r \sin \theta},
\end{equation}
where $v'=L/(r \sin \theta)$.

\begin{figure*}[htb]
  \centering
  \includegraphics[width=1\textwidth]{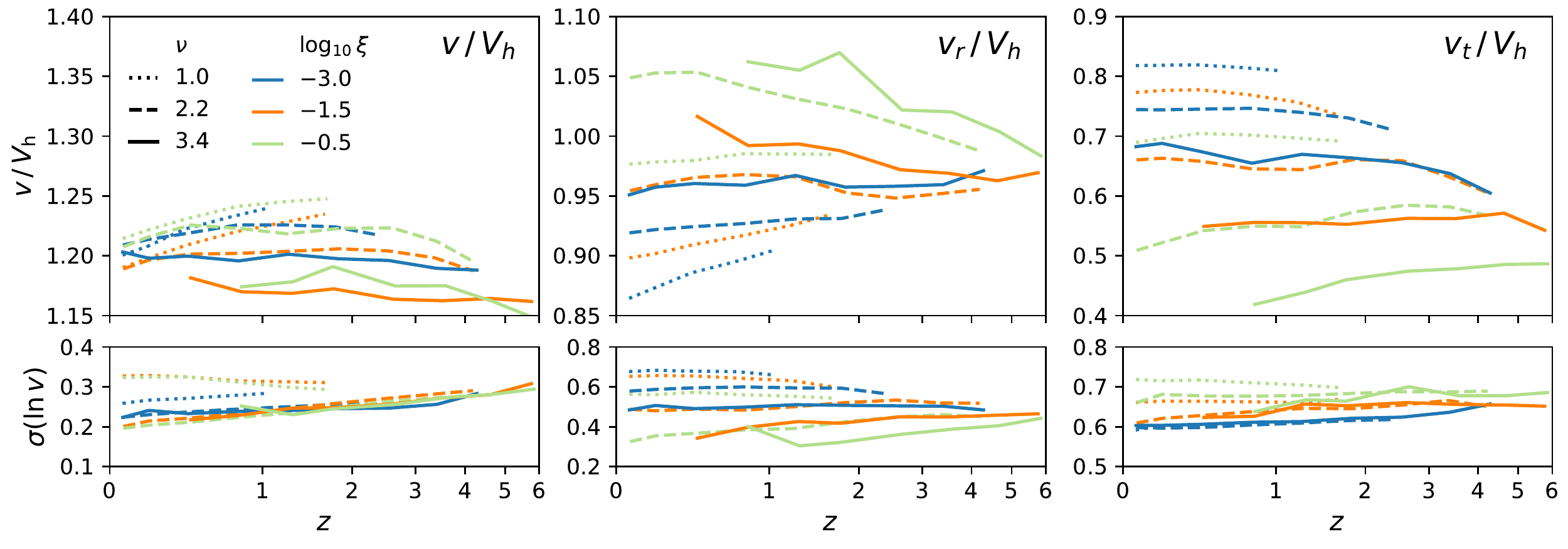}
  \caption{Dependence of subhalo infall velocity on redshift.
  From left to right, the top panels present the median values of $v$, $\vr$ and $\vt$
  as functions of redshift separately.
  The lower panels show the dispersion of the natural logarithm of the corresponding velocities.
  The host mass $\nu$ and sub-to-host ratio $\xi$ used for each line are indicated by the line style and color, respectively.
  The velocity distribution shows no significant evolution with redshift in most cases.
  }
  \label{fig:v_z}
  \vspace{1em}
\end{figure*}

\subsection{Phase-space density of infalling subhalos} \label{sec:phasespace}

Here we derive the relation between the orbital distribution of accreted subhalos in given time interval
and the 6D phase-space density of the infalling subhalos in radial interval near the halo radius.
While the former is directly related to halo assembly as mentioned in \refsec{sec:sample},
the latter is more relevant to the underlying dynamics
and helpful in understanding the model in \refsec{sec:model}.

The phase-space density is defined as $f(\bm{r}, \bm{v})=\dif^6 N / \dif^3 \bm{r} \dif^3 \bm{v}$.
Because $f(\bm{r}, \bm{v})$ only depends on $(r, \vr, \vt)$ under spherical symmetry,
we can write $f(\bm{r}, \bm{v}) = f(r, \vr, \vt)$ or $f(r, v, \theta)$
for convenience, though $f$ still represents the 6D phase-space density (see e.g., \citealt{Li2019a}).
We only consider the distribution at the virial radius $r=\rh$,
then $f$ (as function of $\bm{v}$) actually stands for the velocity ellipsoid.
Note that here $f$ represents the distribution of the subhalos at infall time,
which is different from the present-day distribution.
One can obtain the latter by integrating the former over the halo assembly history
with necessary treatments of subhalo disruption and orbital evolution.

For fixed $v$ and $\theta$, the subhalos in a shell $r\in [\rh, \rh+\vr \dif t]$
can enter the halo (and hence our sample) during the time interval $\dif t$.
The volume of the shell is then $\dif^3\bm{r}=4 \pi \rh^{2}\, v \cos \theta \dif t$.
Now we can calculate the number of such infalling subhalos during $\dif t$,
\begin{align}
  p (v, \theta) \dif v \dif \theta 
  & \propto  \int_{\bm{r},\phi} f (\bm{r}, \bm{v}) v^2 \sin \theta \dif v \dif \theta \dif \phi \dif^3\bm{r} \nonumber\\
  & =  8 \pi^2 \rh^2 \dif t \times v^3 \sin \theta \cos \theta f (\rh, v, \theta) \dif v \dif \theta \nonumber\\
  & \propto  v^3 f (\rh, v, \theta) \dif v \dif \cos^2 \theta.
\end{align}
So the orbital distribution of accreted subhalos, $p (v, \theta)$, differs
from the phase-space distribution, $f(r\!=\!\rh, v, \theta)$, by 
a factor of $v^3 \sin \theta \cos \theta$ besides the normalization.

If the subhalos move \textit{isotropically}
such that $f(\bm{r}, \bm{v})$ is independent of $\theta$,
then $\cos^2 \theta$ of infalling subhalos follows a uniform distribution
that corresponds to $\eta=0$ in \refeqn{eq:theta_distr}.

Moreover, it can be seen that the typical velocity of the accreted subhalos
is larger than the characteristic velocity in phase space because of the factor $v^3$.
Interestingly, solving $\mathrm{argmax}_{v,\theta} f (r\!=\!\rh, v, \theta)$
we find that the most probable velocity vector in the phase space is
$v_\mathrm{mode}=\mu_1 \me^{-4\sigma_1^2} \vh \simeq \vh$ and $\theta_\mathrm{mode}=0$
(i.e., $\vr\simeq\vh,\ \vt=0$),
which is exactly the expectation of spherical collapse.
This is clearly shown later in \reffig{fig:phasespace}.

\section{Model validation} \label{sec:result}

In this section,
we validate the model of the subhalo initial orbital distribution using cosmological simulation data.
We first examine the general trends of the mass and redshift dependence in \refsec{sec:z},
then study the detailed distribution of velocity and infall angle, respectively, in \refsec{sec:vel} and \ref{sec:theta},
and the joint distribution in \refsec{sec:joint}.
The interpretation and discussion of the results are given in \refsec{sec:discussion}.

\subsection{General trends} \label{sec:z}

\reffig{fig:v_z} shows the median and dispersion of the infall velocity, $v$,
and the radial and tangential components, $\vr$ and $\vt$,
as functions of redshift respectively.
The time evolution of the median velocities is generally smaller than $5\%$
except for the major mergers of the largest halos at high $z$.
For the first time,
we demonstrate that the orbital distribution of infalling subhalos is nearly independent of redshift
when the host peak height, $\nu$, and sub-to-host ratio, $\xi$, are controlled.

\begin{figure}[htbp]
  \centering
  \includegraphics[width=0.48\textwidth]{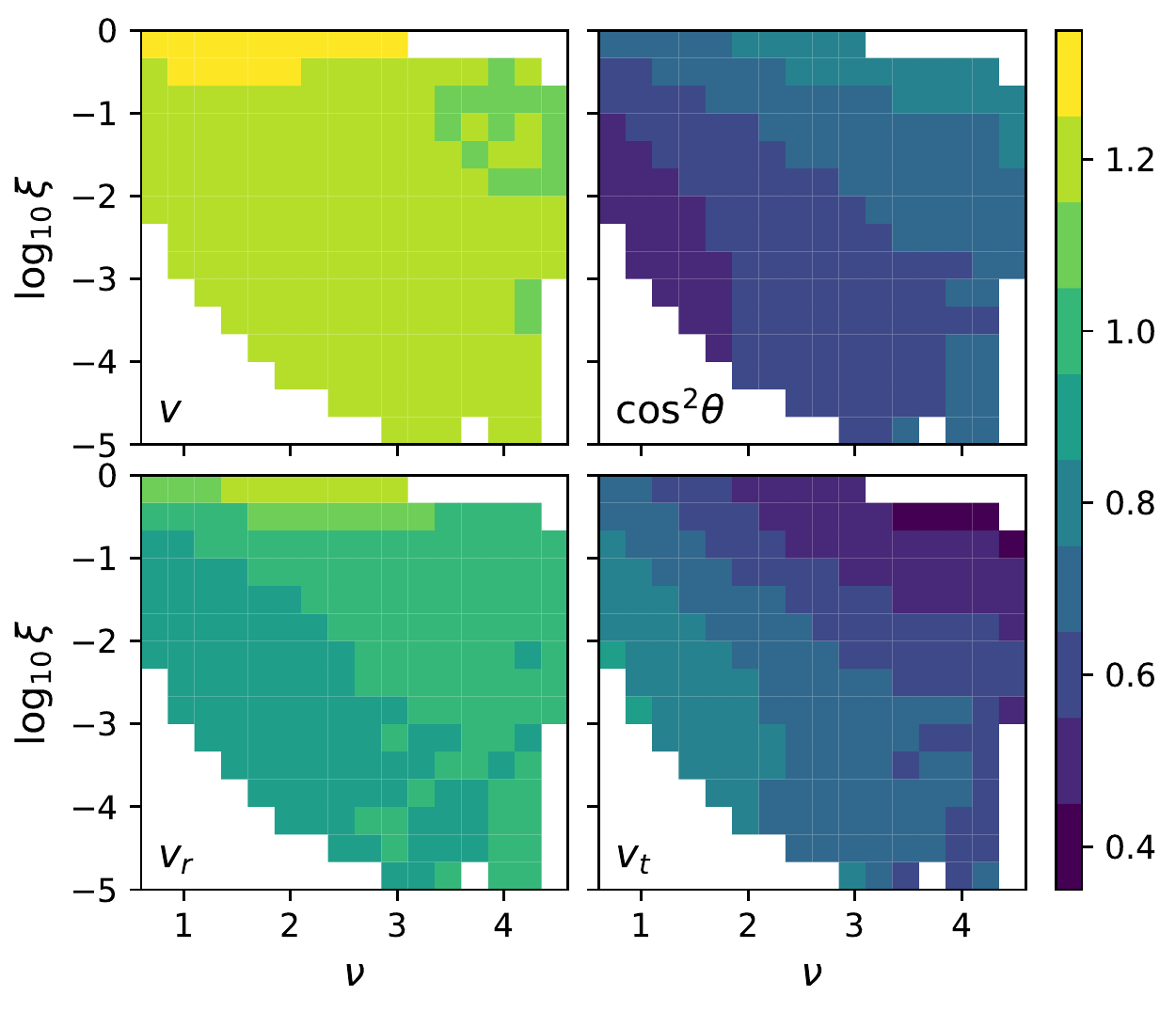}
  \caption{Orbital parameters as a function of the host mass $\nu$ and sub-to-host ratio $\xi$.
  The four panels show respectively the median values of $v,\ \vr$, and $\vt$
  and the mean of $\cossq=\vr^2/v^2$
  for merging halos binned by $\nu$ and $\xi$.
  All velocities are shown in units of the host virial velocity.
  The subhalos with higher $\nu$ or $\xi$
  are more likely to fall along the radial direction due to smaller $\vt$,
  though they have a similar $\vr$ and almost the same total velocity 
  as their low $\nu$, $\xi$ counterparts.
  }
  \label{fig:v_nu_xi)}
\end{figure}

On the other hand, the orbits show apparent systematic changes with both $\nu$ and $\xi$.
The mass dependence is illustrated more clearly in \reffig{fig:v_nu_xi)},
where we show the halos within the whole redshift range with better statistics.
The subhalos with larger host halo or larger sub-to-host ratio
have smaller $\vt$ and slightly larger $\vr$, hence more radial orbits,
while their total velocities keep almost the same median value $\sim 1.2\ \vh$ 
(with variation $\lesssim 5\%$; see also the left panel of \reffig{fig:v_z}).
Our result confirms the dependence on the mass of both halos reported previously \citep{Tormen1997,Wang2005,Jiang2015}
but in a larger dynamic range,
especially the very massive cluster halos and major mergers
that were generally poorly covered.

\citet{Wetzel2011} found that for halo pairs of fixed masses,
$\avg{v} \sim 1.15 \vh$ is almost irrelevant to redshift,
while $\vr$ is larger and $\vt$ is smaller at higher $z$.
This trend is consistent with our findings,
noting that $\nu$ for fixed mass is larger at higher $z$.
\citeauthor{Wetzel2011} has also examined that 
picking halos with fixed $\mh/M_\ast(z)$ instead of $\mh$ at different redshifts
cannot remove the redshift dependence,
where $M_\ast(z)$ is the characteristic mass (s.t.\ $\nu(M_\ast, z)=1$).
He interpreted it as an intrinsic redshift dependence,
which is contrary to our findings using $\nu$.
The discrepancy is because 
the matter power spectrum of the Universe is not scale-free,
so that $\nu$ is a better representative than $M/M_\ast$ to characterize the halo size across cosmic time.

\subsection{Velocity} \label{sec:vel}

As shown in \reffig{fig:v_z}, the median and the dispersion of the 
velocity $v$ are almost independent of redshift and mass,
which implies a nearly universal distribution $p(v)$.

\begin{figure}[tbp]
  \centering
  \includegraphics[width=0.48\textwidth]{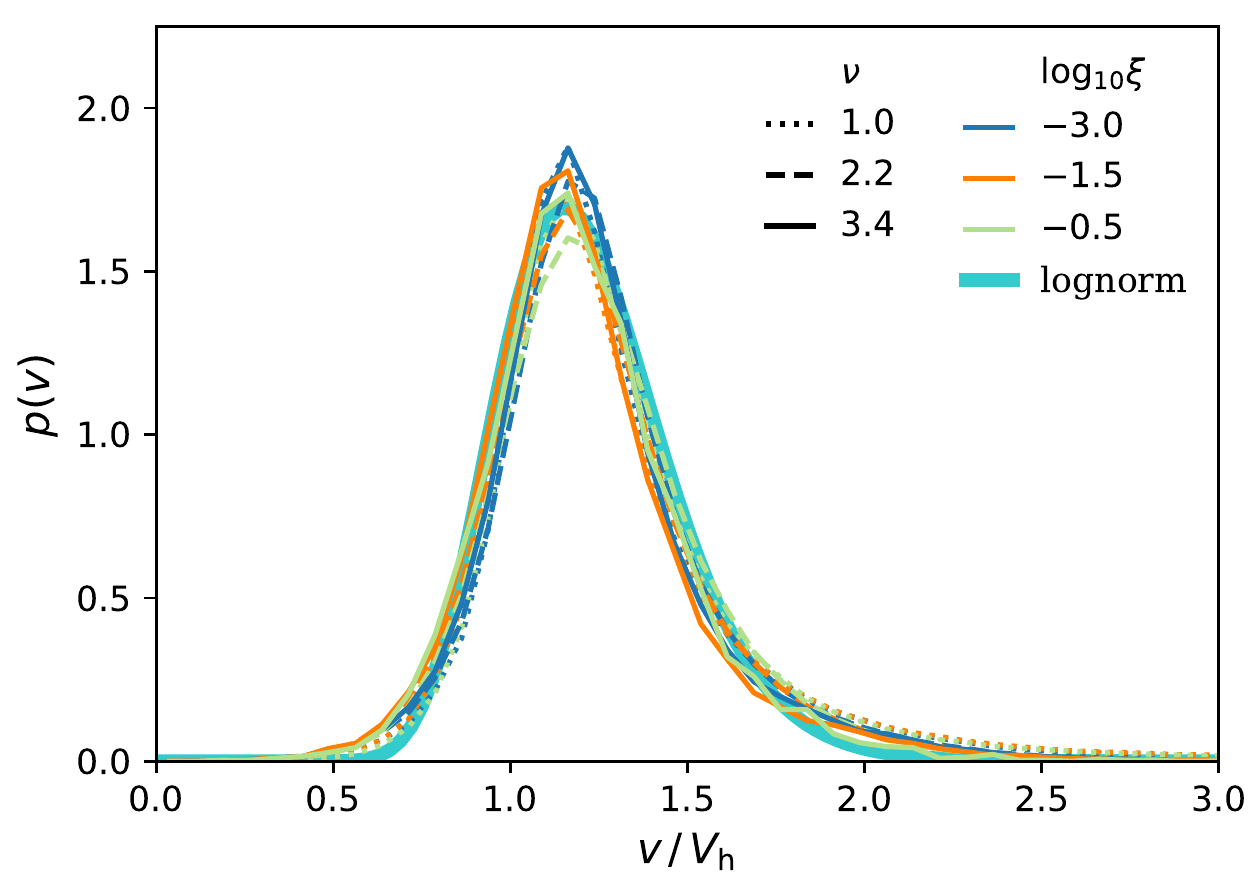}
  \caption{Infall velocity distribution for different host mass $\nu$ and sub-to-host ratio $\xi$ bins.
  $\nu$ and $\xi$ of each bin are indicated by the line style and color, respectively.
  The cyan thick solid line shows our model (\refeqnalt{eq:v_distr}), which is a log-normal distribution.
  }
  \label{fig:p_v}
\end{figure}

The independence on redshift allows us to use a larger sample by combining data at different epochs
to study the mass dependence in detail.
\reffig{fig:p_v} shows the velocity distributions 
of subhalos with $z<4$ in various host mass $\nu$ and sub-to-host ratio $\xi$ bins.
We find that the log-normal distribution (\refeqnalt{eq:v_distr})
presents a good description for all cases. 
The distribution peaks at $\mu_1\me^{-\sigma_1^2}\simeq 1.15$, 
which has been reported in various literature \citep{Wetzel2011,Jiang2015}.

However, looking closely into the figure,
the data shows more extended tails than log-normal, especially for $v>1.7\vh$.
It is possible to find a more complex form to fit the long tail of the distribution.
For now, we are satisfied with log-normal for its simplicity,
also because the subhalos with extremely high velocity are likely to be fly-bys,
thus less interesting for semi-analytical models.
Nevertheless, it is worth understanding better the origin and possible influence of such high-speed subhalos in the future.

Moreover, one could find a weak mass dependence for velocity distribution.
In particular, the subhalos within small host halos ($\nu=1$, dotted lines)
show more extended tails at the high-velocity end ($v>2\vh$) in \reffig{fig:p_v}
and, consequently, have larger velocity dispersions as shown in the lower panel of \reffig{fig:v_z}.
As we argue in \refsec{sec:environ}, this is likely because the environment 
is relatively hotter for smaller halos. 
Their satellites might get accelerated greatly by the external tidal field 
from massive neighboring clusters and large-scale structures.

There are other functional forms used to fit the velocity distribution,
e.g., the Maxwell distribution \citep{Vitvitska2002} and Voigt distribution \citep{Jiang2015},
which, however, cannot reveal the skewed tail at large $v$ as log-normal.

\subsection{Infall angle} \label{sec:theta}

\begin{figure*}[!htbp]
  \centering
  \includegraphics[width=0.85\textwidth]{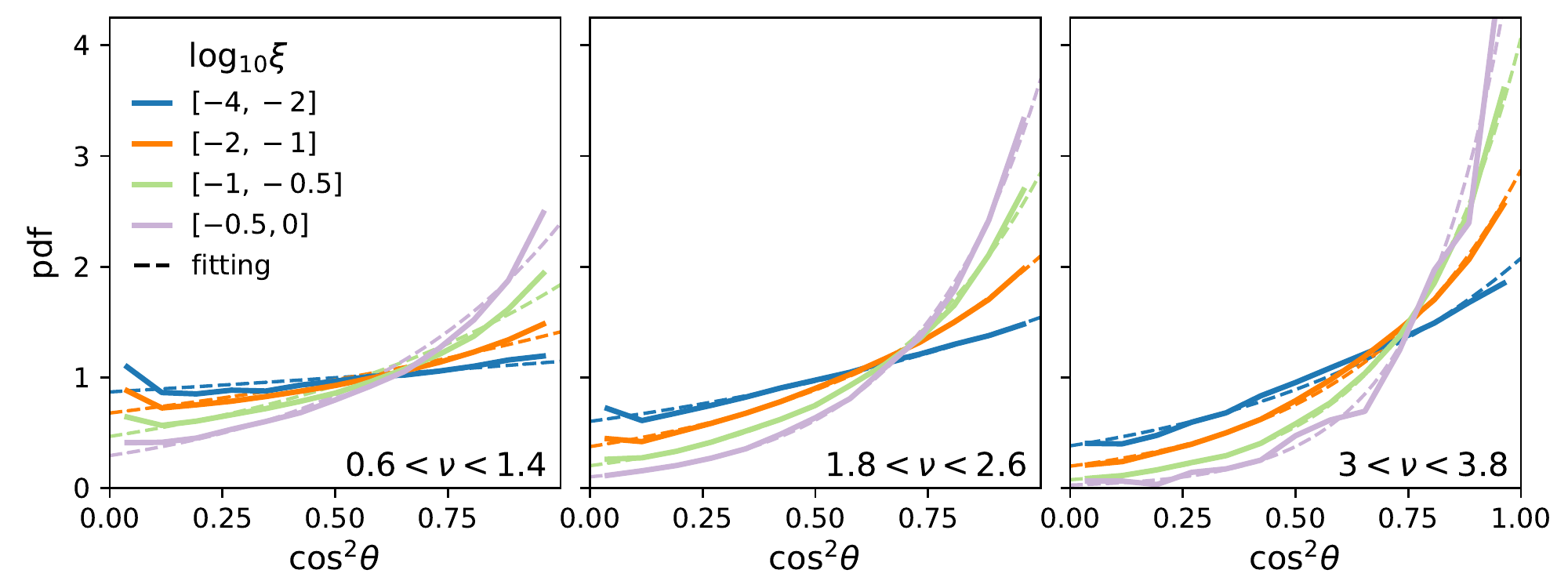}
  \caption{Distribution of the infall angle in terms of $\cossq$ for 
  different host mass $\nu$ (panels) and sub-to-host ratio $\xi$ (colors) bins.
  The solid lines show the distributions of our halo sample, 
  while the dashed lines represent fits of exponential function 
  characterized by a single parameter $\eta$ (\refeqnalt{eq:theta_distr}).
  }
  \label{fig:theta_nu}
\end{figure*}

\begin{figure*}[!htbp]
  \centering
  \includegraphics[width=0.85\textwidth]{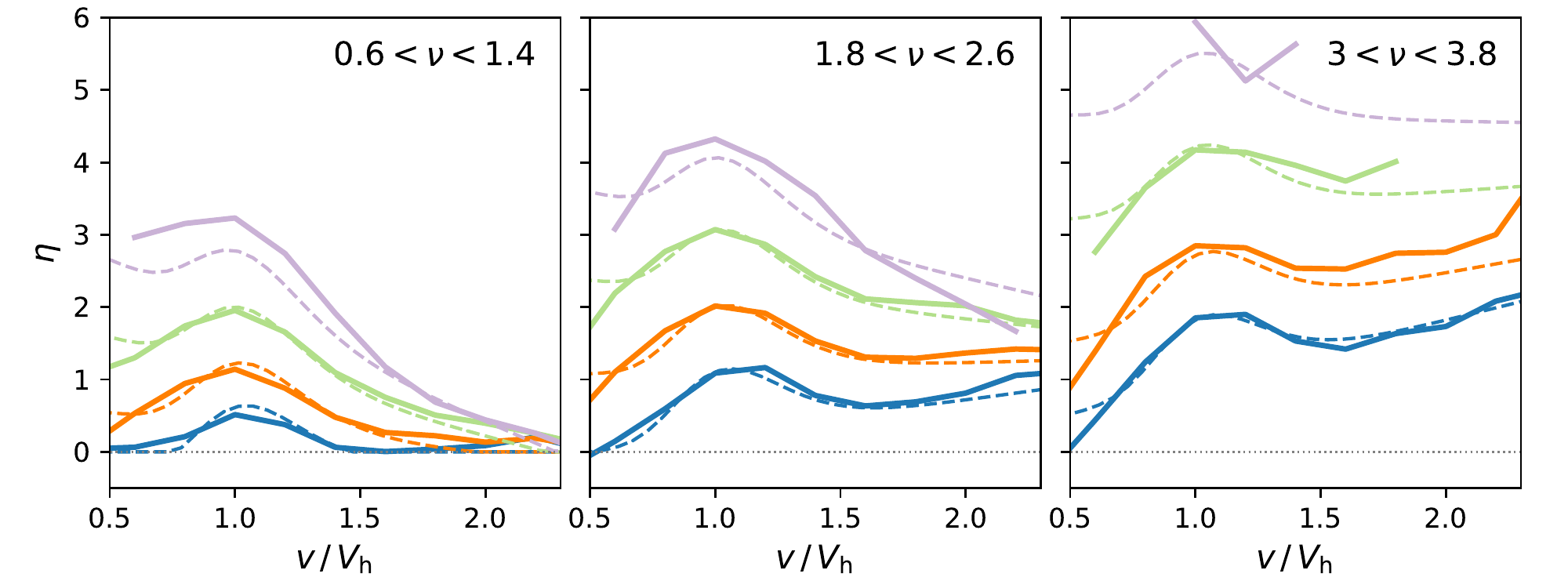}
  \caption{Distribution of infall angle as function of velocity.
  The distribution of angle is fully characterized by single parameter $\eta$
  (\refeqnalt{eq:theta_distr}), which depends on the velocity $v$ ($x$-axis), 
  the host mass $\nu$ (panels) and sub-to-host ratio $\xi$ (colors; see the legend of \reffig{fig:theta_nu}).
  The solid lines show the value of $\eta$ fitted individually in each $(v,\nu,\xi)$ bin;
  the dashed lines show our unified model (\refeqnalt{eq:eta}).
  Note that larger $\eta$ stands for more radial infall direction.
  }
  \label{fig:eta_nu}
  \vspace{1em}
\end{figure*}

\reffig{fig:theta_nu} shows the distribution of the infall angle 
in terms of $\cossq=\vr^2/v^2$ for different $\nu$ and $\xi$ bins.
$\cossq$ can be perfectly described by the exponential function (\refeqnalt{eq:theta_distr}), 
$\dif{N}/\dif{\cossq} \propto \exp \left({\eta \cossq} \right)$.
We use the relation $\avg{\cossq} = \me^\eta/(\me^\eta-1) - 1/\eta$ to solve the best-fit $\eta$
from the average of $\cossq$ in each bin.
Note that a larger $\eta$ implies that $\theta$ is smaller on average and the orbits are more radial,
while $\eta=0$ represents isotropic inflow pattern (see \refsec{sec:phasespace}).
The mass dependence of the infall angle is again confirmed and consistent 
with the trends in \refsec{sec:z} that $\theta$ is smaller
for larger $\nu$ or $\xi$.

A complete description of the orbits comprises the velocity distribution, $p(v)$,
and the conditional distribution of the falling angle for given velocity, $p(\theta|v)$.
Due to the limitation of sample size in the previous work, 
$p(\theta|v)$ has not been investigated directly.
\citet{Jiang2015} assumed that $p(\theta)$ is independent of $v$.
However, they did not examine it explicitly.
On the contrary, we find that $p(\theta)$ actually varies with $v$ in a non-monotonic way.

We find that the exponential function is still an excellent description 
for $p(\cossq)$ if further binning the sample by $v$,
so we can use a single parameter $\eta$ as representative of $p(\cossq)$.
\reffig{fig:eta_nu} shows $\eta$ as function of $v$.
When $v$ is close to $\vh$, the virial velocity of the host, 
$\eta$ reaches its maximum so that the subhalo orbits are most radially distributed.
When $v$ is much smaller or larger than $\vh$, the subhalos tend move more isotropically.
Our model prediction from \refeqn{eq:eta} and \reftab{tab:orbit_param}
well captures the general trends, especially for $v\sim \vh$.

It is worth pointing out that $\eta \geqslant 0$ for nearly all the $(v, \nu, \xi)$ bins
(also justified by $\avg{{\cossq}} \geqslant 0.5$ in \reffig{fig:v_nu_xi)}),
which means the radial motion is always stronger than or at least equal to the tangential one
in the sense of $\avg{\vr^2/v^2} \geqslant \avg{\vt^2/v^2}$.

Finally, similar but non-equivalent exponential forms have been proposed to describe the angle distribution
by \citet{Wang2005} and \citet{Jiang2015}.
Nevertheless, we find our \refeqn{eq:theta_distr} provides a more accurate description to the data shown in \reffig{fig:theta_nu}.

\subsection{Joint distribution} \label{sec:joint}

\begin{figure*}[htbp]
  \centering
  \includegraphics[width=0.498\textwidth]{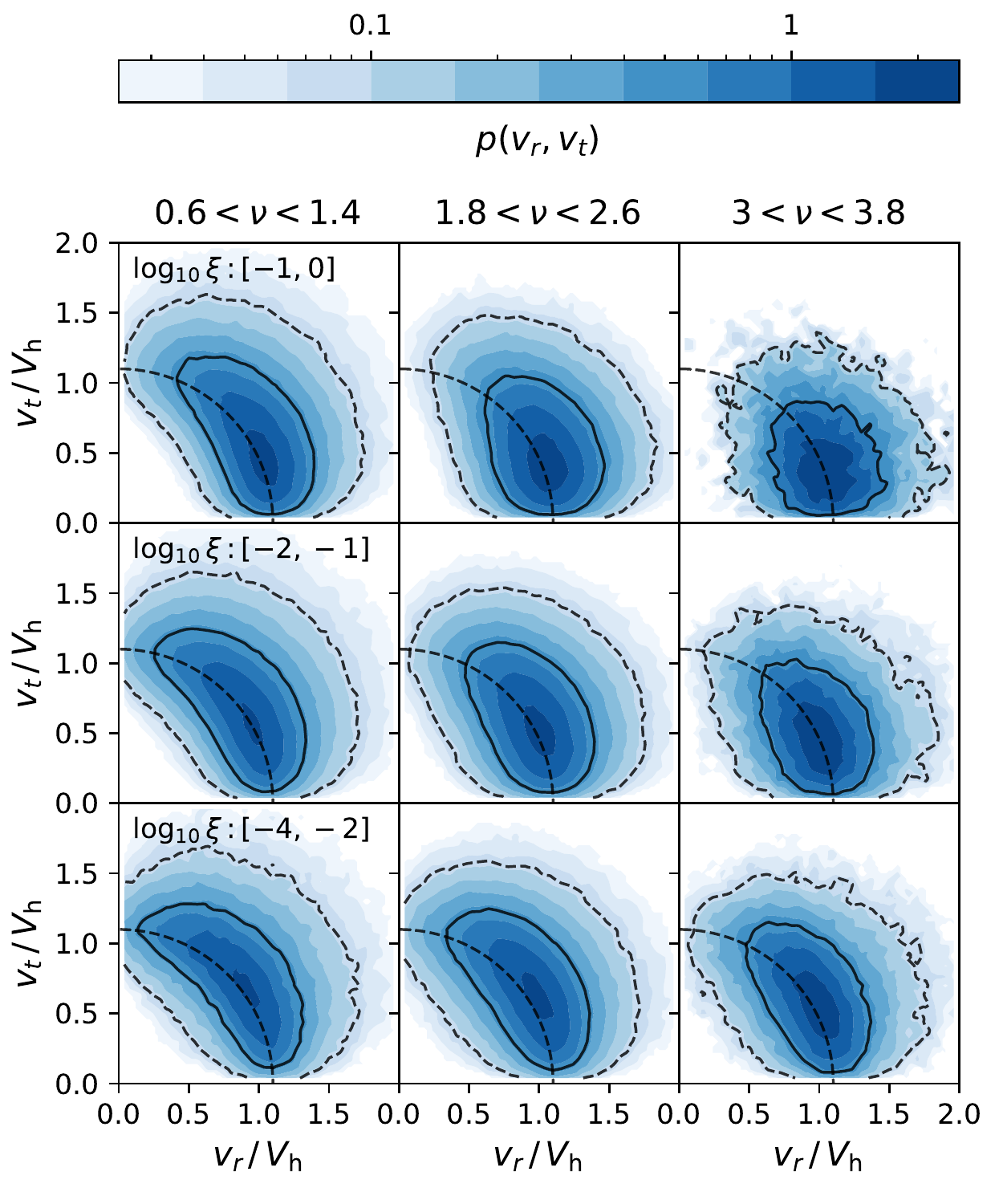}
  \includegraphics[width=0.498\textwidth]{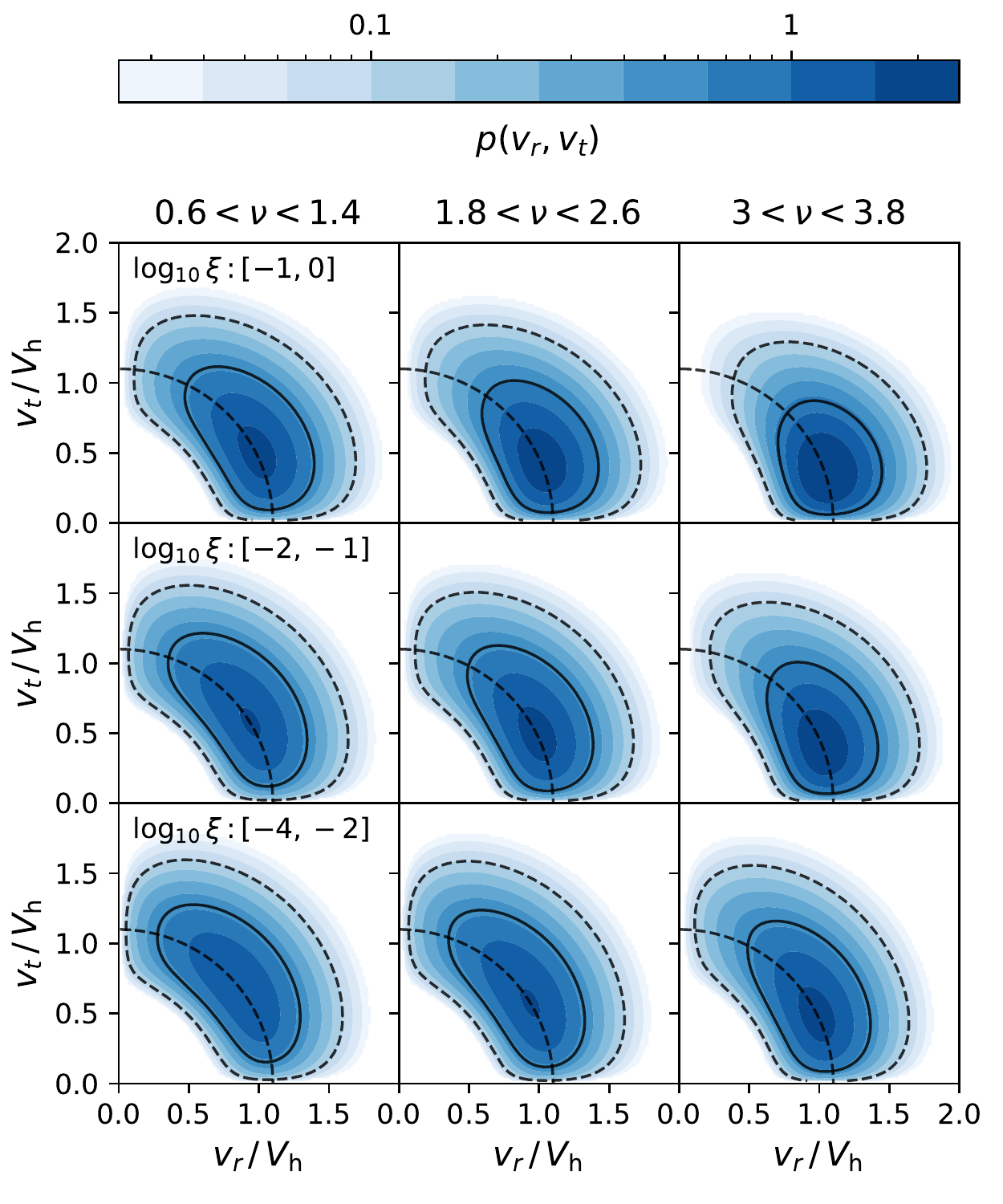}
  \caption{$p(\vr, \vt)$ of infalling subhalos (left panels)
  and our model prediction (right panels)
  for various host mass $\nu$ (columns) and sub-to-host ratio $\xi$ (rows).
  The colored contours show the probability density using the same color normalization.
  The blue solid and dashed contour lines show the 68\% and 95\% regions respectively,
  while the black dashed arcs show $\vr^2+\vt^2=(1.1\vh)^2$ for reference.
  }
  \label{fig:p_vr_vt}
\end{figure*}

\begin{figure}[htbp]
  \centering
  \includegraphics[width=0.498\textwidth]{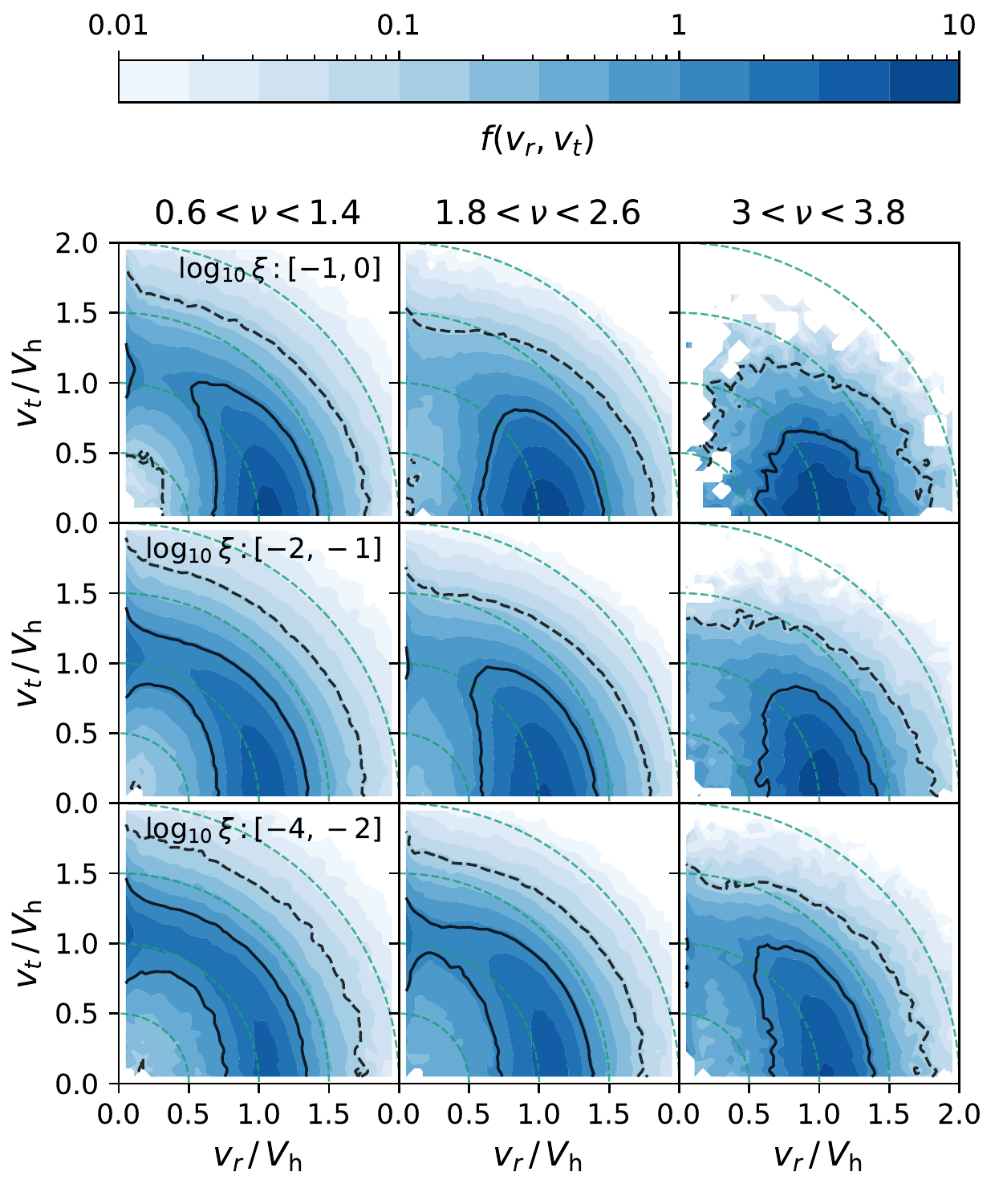}\\
  \caption{Similar to left panel of \reffig{fig:p_vr_vt}, 
  but showing the phase-space density derived in \refsec{sec:phasespace} for the simulation sample.
  The colored contours represent $f(\rh, \vr,\vt)$ as function of $\vr$ and $\vt$, 
  which is actually a \textit{slice of the velocity ellipsoid}.
  It is clear that the velocity always peaks at $\vh$.
  The velocity ellipsoid approaches isotropic when $\nu$ and $\xi$ are both small.
  }
  \label{fig:phasespace}
\end{figure}

We further demonstrate the validity of our model with the 2D velocity distribution $p(\vr,\vt)$
in \reffig{fig:p_vr_vt}.
The right panels show our model prediction using the variable transformation in \refeqn{eq:vr_vt}.
Our model well captures the shapes of these distributions 
and mass dependence for various $\nu$ and $\xi$ bins.
Despite the clear mass dependence, 
the general similarity of the velocity distributions is also notable,
especially when the sub-to-host ratio is small (as for most subhalos).

\reffig{fig:phasespace} shows the phase-space density
of infalling subhalos derived in \refsec{sec:phasespace}.
The contours represent $f(\rh,\,\vr,\,\vt)$ as a function of $\vr$ and $\vt$,
which is actually a \textit{slice of the velocity ellipsoid}.
The complete velocity ellipsoid, i.e., the probability density distribution in the 3D velocity space,
can be obtained by rotating each panel around the $x$-axis.
It is clear that the velocity ellipsoid 
always peaks at $\vr\simeq\vh$ and $\vt=0$, which is exactly the expectation of spherical collapse.
The velocity ellipsoid shows a strong radial inflow pattern when $\nu$ or $\xi$ is large,
while it approaches isotropic when $\nu$ and $\xi$ are both small.
Moreover, the contours also show that
the phase-space distribution is more uniform along $\theta$
when $v$ is significantly larger or smaller than $\vh$.

\section{Discussion} \label{sec:discussion}

\subsection{Interpretation of results from environmental effects} \label{sec:environ}

The external tidal field raised from nearby objects and the large scale structure
can change the velocity and moving direction of infalling subhalos,
therefore, leading to deviation from the idealized radial inflow with $\vh$ expected by spherical collapse.
Consequently, halos have different accretion flow patterns in different environments
\citep[e.g.,][]{Shi2015,Kang2015a,Wang2017a} and hence different internal properties \citep{Wang2011,Shi2015,Chen2016}.
In particular, the motions of subhalos tend to align with the tidal field's principal axes \citep{Shi2015}.
In the extreme case, if the local potential is completely dominated by the external field,
the velocity ellipsoid of subhalos should be isotropic on average, 
because the radial direction related to the target halo is no longer special.
It is exactly what we have seen for the case that $\nu$ and $\xi$ are both small in \reffig{fig:phasespace}.
We believe the results reported in this paper can be understood intuitively by environmental factors as follows.

\textbullet~ 
Mass dependence.
It is obvious that the relative strength of external effects depends on the halo mass.  
Very massive halos generally dominate the surrounding potential field, 
so naturally, their satellites fall rather radially with $v \sim \vh$.
It is also consistent with the halo formation theory,
that more massive halos form from more spherical overdensities
with accreting matter containing less specific angular momentum \citep{Zeldovich1970,Bardeen1986,Sheth2001}.
Perhaps similarly, more massive subhalos are more resistant
to perturbations from nearby subhalos
and hence able to better keep their infalling motions.
Conversely, low mass halos and low mass subhalos are more vulnerable 
to environmental factors that lead to a more random orbital direction
and hence a larger tangential component.
Interestingly, 
we can only observe isotropic inflow when $\nu$ and $\xi$ are both small.
Even for the smallest halos in our sample,
the major mergers still shows a kind of radial inflow pattern.
Therefore, the dependence of $\nu$ and that of $\xi$ 
seem to be indeed representations of different aspects of the environment.

\textbullet~ 
$p(v)$ and $p(\theta|v)$.
The external field can either accelerate or decelerate the velocity components
of a subhalo depending on its relative moving direction.
Eventually, it likely affects the direction of motion more than the amplitude of velocity on average.
It is probably why we find a nearly universal $p(v)$ but an angular distribution depending on mass.
Nevertheless, low mass hosts indeed have slightly more subhalos with extreme velocities,
as shown in \reffig{fig:p_v}.
Moreover, any significant deviation from the ideal velocity of $\vh$ 
implies strong external effects typically associated with random motions.
This explains the correlation between $v$ and $\theta$
that subhalos have more radial orbits when $v\sim\vh$, and more random directions otherwise (\reffig{fig:eta_nu}).

Finally, if the mass dependence is a reflection of the environmental dependence, as argued above, 
then it is at most a partial reflection,
because halos of the same masses can reside in very different environments.
For example, small halos can also dominate locally 
and have similarly strong radial inflow as massive halos
if they are rather isolated.
The final state of a halo should rely on
the whole history of both mass growth and the environment.

\subsection{Dependence on cosmology} \label{sec:cosmo}

As shown in \refsec{sec:z},
the subhalo orbital distribution is approximately
independent of redshift when the host peak height, $\nu$,
and the mass ratio, $\xi$, are controlled.
Noting the dramatic changes of the cosmological parameters
across cosmic time in our simulations (e.g., 
$\Omega_\mathrm{m}(z)$ evolves from $0.99$ to $0.27$
and $\sigma(8\, \mpch, z)$, the amplitude of the matter power spectrum at $8\, \mpch$,
evolves from 0.16 to 0.85 since $z=6$),
it seems reasonable to speculate that our results 
are possibly insensitive to the cosmology, 
especially to the cold dark matter (CDM) cosmologies.
It is because the dimensionless variables,
$\nu$ and $\xi$, that are motivated by halo formation theory,
present the relative rank of the halos and subhalos 
in a very general way.
This virtue has also been justified by the 
approximately universal halo virial mass function
(e.g., \citealt{Despali2016,Diemer2020}),
halo growth history (e.g., \citealt{Zhao2009,vandenBosch2014}),
and subhalo mass function
(e.g., \citealt{Gao2004,Han2018a})
in terms of $\nu$ or $\xi$.

In other cosmologies,
e.g., the warm dark matter (WDM) model,
the mass function and internal structure of halos or subhalos
are very different from those in CDM
due to the different shape of matter power spectrum.
We speculate that our results (or at least the trends)
might still hold to a certain extent,
because the inflow of subhalos is mainly dominated by the gravity of their host after all.

However, it is important to note that 
all the simulations used in this work have the same cosmology;
therefore, here we are not able to justify the above arguments directly.
In the future, the pertinent simulations of different cosmologies
should be used to quantify the possible cosmology dependence.

\section{Conclusion} \label{sec:conclusion}

In this paper, we are aiming to provide a comprehensive and unified description of 
the initial orbits of infalling subhalos
\textit{across cosmic time}.
Using 16 cosmological simulations of various resolutions,
the unprecedented large sample size and dynamic range allow us to
characterize the joint distribution of orbital parameters and 
its dependence on the host mass (in terms of peak height), $\nu$ , 
sub-to-host mass ratio, $\xi=m/\mh$, and redshift, $z$, with high precision.

An accurate but simple model, $p(v,\theta|\nu, \xi; z)$, is proposed (\refsec{sec:model}) and 
validated with simulations (\refsec{sec:result}). 
More specifically, we find that:
\begin{asparaitem}
\item 
The infall velocity, $v$, follows a nearly universal 
log-normal distribution
(fig.~\ref{fig:p_v}),
so does the orbital energy.
The most probable velocity vector of an infalling subhalo in the phase space locates
at $v_r\simeq\vh$ and $v_t=0$ as expected by the spherical collapse model
(fig.~\ref{fig:phasespace}).

\item 
The infall angle in terms of $\cossq=\vr^2/v^2$ follows an exponential distribution
that depends on $v$, $\nu$ and $\xi$
(fig.~\ref{fig:theta_nu}).
The orbits are most radially biased when $v\sim\vh$ 
(fig.~\ref{fig:eta_nu}).

\item 
Subhalos with higher host mass, $\nu$, or higher sub-to-host ratio, $\xi$, have more radial orbits
with relatively smaller angular momentum or pericenter distance
(fig.~\ref{fig:v_nu_xi)}).
Subhalos tend to move isotropically when $\nu$ and $\xi$ are both small
(fig.~\ref{fig:phasespace}).

\item 
The above description is nearly independent of redshift (fig.~\ref{fig:v_z}) 
and possibly insensitive to cosmology (sec.~\ref{sec:cosmo}).

\item 
It is consistent with the scenario where the dynamical environment is relatively colder for massive structures
because their gravity more likely dominates the local potential
(sec.~\ref{sec:environ}).

\item 
The approximately universal velocity distribution and the mass-dependent angle distribution seem to imply that
the external tidal fields generally affect the direction rather than the amplitude of subhalo velocity on average.
\end{asparaitem}

We have confirmed the mass dependence of subhalo orbits reported in the literature (see \refsec{sec:intro})
in a much larger dynamic range with better statistics.
More importantly, we have proposed a unified quantitative description validated across cosmic time.

Note that our data cover $0.6 \leq \nu \lesssim 4.5$ 
(corresponding to $10^{11.2\sim15.6}\msunh$ at $z=0$ and $10^{1.7\sim12.2}\msunh$ at $z=6$; 
see \reffig{fig:n_nu_xi}),
$10^{-5} \lesssim \xi \leq 1$, and $0 \leq z \lesssim 6$;
it is remarkable that a simple model can well characterize the orbital distribution
for such wide parameter space.
The simplicity and accuracy of our model could be attributed to 
the use of the dimensionless variables ($\nu$ and $\xi$) motivated by halo formation theory
and the appropriate separation of the physical components.
It warrants further investigation of the mechanism behind.
In addition, future work should examine the model in a larger dynamical range
(especially the halos of $\nu < 0.6$)
and quantify the possible cosmology dependence with pertinent simulations.

Our model can be used as the initial condition in semi-analytic models of galaxy formation
(e.g., \citealt{Yang2011,Jiang2020}; S. Green et al.\ in prep.\ 2020) along with the halo growth history \citep{Zhao2009} 
and subhalo mass function (e.g., \citealt{Gao2004,Han2018a}) 
and accretion rate (e.g., \citealt{Lacey1993,Yang2011,Fakhouri2010}; F.Y.\ Dong et al.\ in prep.).
It also enables a better understanding of the halo structures and their dependence on environments.
For example, 
more massive halos are expected to have higher velocity anisotropy (see, e.g., \citealt{Lemze2012}),
so are the isolated halos.
Moreover,
the final dynamical state of a halo should depend on the whole history
considering that $\nu$ can change with time and result in different accretion patterns.

Besides the mass dependence discussed above,
we have to emphasize the general similarity in subhalo infall pattern across cosmic time,
especially when the sub-to-host ratio is small (as for most of the subhalos and probably the diffuse matter).
Subhalos are the building blocks of halos;
therefore, this similarity in initial kinematics and the universal subhalo mass function 
might eventually help to understand
the many reported universal self-similar halo properties, 
e.g., the density profile \citep{Navarro1996,Navarro2004},
pseudo-phase-space profile \citep{Taylor2001,Navarro2010},
angular momentum distribution \citep{White1984,Bullock2001b},
and the subhalo spatial distribution \citep{Gao2004,Springel2008,Jiang2016,Han2016} and 
kinematics \citep{Li2017,Li2019a}.

Finally, we have only considered the average subhalo orbital distribution under the spherical symmetry in this work.
A more comprehensive description of subhalo kinematics can further
integrate the group infall and anisotropic accretion of subhalos \citep[e.g.,][]{Benson2005,Libeskind2014,Shi2015,Kang2015a,Shao2018a}, 
which are crucial for understanding
the alignment among the halo shape, spin, and the large-scale structure \citep[e.g.,][]{Wang2011,Chen2016,Wang2017a,Morinaga2020}
and interpreting the anisotropic distribution of satellite galaxies reported in 
observations \citep[e.g.,][]{Kroupa2005,Yang2006,Pawlowski2012,Ibata2013,Wang2020a}.


\vspace{1cm}
We are very grateful to Chunyan Jiang, Fangzhou Jiang, Xianguang Meng, Houjun Mo, Yongzhong Qian, 
Feng Shi, Jingjing Shi, and Xiaohu Yang for their helpful discussions,
and to Frank C. van den Bosch, Sheridan B. Green, and Peng Wang for careful reading of the manuscript and insightful comments.
We also thank the anonymous referee for constructive criticisms and helpful suggestions.
This work is supported by NSFC (11222325, 11533006, 11621303, 11873038, 11890691, 11973032), 
the Knowledge Innovation Program of CAS (KJCX2-EW-J01),
Shanghai talent development funding (2011069),
National Key Basic Research and Development Program of China (No.\ 2018YFA0404504),
and the 111 project (No.\ B20019). 
We gratefully acknowledge the support of the Key Laboratory for Particle Physics, 
Astrophysics and Cosmology, Ministry of Education.

This work made use of the High Performance Computing Resource 
in the Core Facility for Advanced Research Computing at Shanghai Astronomical Observatory.

\textit{Software:} 
  HBT subhalo finder and tree maker \citep{Han2012},
  Colossus \citep{Diemer2017a},
  Astropy \citep{AstropyCollaboration2013},
  Numpy \citep{Walt2011}, 
  Scipy \citep{Oliphant2007},
  Matplotlib \citep{Hunter2007}

\appendix \label{appendix}

\section{Orbit Interpolation} \label{sec:interp}
We use the cubic spline to interpolate the subhalo orbits between adjoint snapshots
to find the precise crossing time.
Cubic splines are smooth curves with continuous second-order derivative (aka the acceleration).
Specifically, each component of the position and velocity $\{x_i, v_i\}_{i=1,2,3}$
is interpolated separately by the following equation:
\begin{equation} \begin{aligned}
    x_i(t) =& a_{i0} + a_{i1} t + a_{i2} t^2 + a_{i3} t^3, \\
    v_i(t) =& a_{i1} + 2 a_{i2} t + 3 a_{i3} t^2. \\
\end{aligned} \end{equation}
The 12 unknowns, $\{a_{ij}\}$, are determined 
by substituting $x_i$ and $v_i$ at adjoint snapshots.
We approximate the halo growth by an exponential function of time between the snapshots,
$\ln \rh(t)=b_0 + b_1 t$,
then solve the exact infall time $t_\mathrm{inf}$ that satisfies $\abs{\bm{x}(t)}=\rh(t)$
and calculate corresponding $\bm{x}(t_\mathrm{inf}),\bm{v}(t_\mathrm{inf})$.
Similarly, we interpolate the masses of both halos exponentially to the infall time.

\bibliography{orbits}

\begin{thebibliography}{}
\expandafter\ifx\csname natexlab\endcsname\relax\def\natexlab#1{#1}\fi
\providecommand{\url}[1]{\href{#1}{#1}}
\providecommand{\dodoi}[1]{doi:~\href{http://doi.org/#1}{\nolinkurl{#1}}}
\providecommand{\doeprint}[1]{\href{http://ascl.net/#1}{\nolinkurl{http://ascl.net/#1}}}
\providecommand{\doarXiv}[1]{\href{https://arxiv.org/abs/#1}{\nolinkurl{https://arxiv.org/abs/#1}}}

\bibitem[{{Astropy Collaboration} {et~al.}(2013){Astropy Collaboration},
  {Robitaille}, {Tollerud}, {Greenfield}, {Droettboom}, {Bray}, {Aldcroft},
  {Davis}, {Ginsburg}, {Price-Whelan}, {Kerzendorf}, {Conley}, {Crighton},
  {Barbary}, {Muna}, {Ferguson}, {Grollier}, {Parikh}, {Nair}, {Unther},
  {Deil}, {Woillez}, {Conseil}, {Kramer}, {Turner}, {Singer}, {Fox}, {Weaver},
  {Zabalza}, {Edwards}, {Azalee Bostroem}, {Burke}, {Casey}, {Crawford},
  {Dencheva}, {Ely}, {Jenness}, {Labrie}, {Lim}, {Pierfederici}, {Pontzen},
  {Ptak}, {Refsdal}, {Servillat}, \& {Streicher}}]{AstropyCollaboration2013}
{Astropy Collaboration}, {Robitaille}, T.~P., {Tollerud}, E.~J., {et~al.} 2013,
  {\href{https://doi.org/10.1051/0004-6361/201322068}{\aap}},
  {\href{https://ui.adsabs.harvard.edu/abs/2013A&A...558A..33A}{558}}, A33

\bibitem[{{Bah{\'e}} {et~al.}(2019){Bah{\'e}}, {Schaye}, {Barnes}, {Dalla
  Vecchia}, {Kay}, {Bower}, {Hoekstra}, {McGee}, \& {Theuns}}]{Bahe2019}
{Bah{\'e}}, Y.~M., {Schaye}, J., {Barnes}, D.~J., {et~al.} 2019,
  {\href{https://doi.org/10.1093/mnras/stz361}{\mnras}},
  {\href{https://ui.adsabs.harvard.edu/abs/2019MNRAS.485.2287B}{485}}, 2287

\bibitem[{{Balogh} {et~al.}(2000){Balogh}, {Navarro}, \& {Morris}}]{Balogh2000}
{Balogh}, M.~L., {Navarro}, J.~F., \& {Morris}, S.~L. 2000,
  {\href{https://doi.org/10.1086/309323}{\apj}},
  {\href{https://ui.adsabs.harvard.edu/abs/2000ApJ...540..113B}{540}}, 113

\bibitem[{{Bardeen} {et~al.}(1986){Bardeen}, {Bond}, {Kaiser}, \&
  {Szalay}}]{Bardeen1986}
{Bardeen}, J.~M., {Bond}, J.~R., {Kaiser}, N., \& {Szalay}, A.~S. 1986,
  {\href{https://doi.org/10.1086/164143}{\apj}},
  {\href{https://ui.adsabs.harvard.edu/abs/1986ApJ...304...15B}{304}}, 15

\bibitem[{{Baugh}(2006)}]{Baugh2006}
{Baugh}, C.~M. 2006,
  {\href{https://doi.org/10.1088/0034-4885/69/12/R02}{RPPh}},
  {\href{https://ui.adsabs.harvard.edu/abs/2006RPPh...69.3101B}{69}}, 3101

\bibitem[{{Behroozi} {et~al.}(2013){Behroozi}, {Wechsler}, \&
  {Wu}}]{Behroozi2013b}
{Behroozi}, P.~S., {Wechsler}, R.~H., \& {Wu}, H.-Y. 2013,
  {\href{https://doi.org/10.1088/0004-637X/762/2/109}{\apj}},
  {\href{https://ui.adsabs.harvard.edu/abs/2013ApJ...762..109B}{762}}, 109

\bibitem[{{Benson} {et~al.}(2020){Benson}, {Behrens}, \& {Lu}}]{Benson2020}
{Benson}, A., {Behrens}, C., \& {Lu}, Y. 2020,
  {\href{https://doi.org/10.1093/mnras/staa1777}{\mnras}},
  {\href{https://ui.adsabs.harvard.edu/abs/2020MNRAS.496.3371B}{496}}, 3371

\bibitem[{{Benson}(2005)}]{Benson2005}
{Benson}, A.~J. 2005,
  {\href{https://doi.org/10.1111/j.1365-2966.2005.08788.x}{\mnras}},
  {\href{https://ui.adsabs.harvard.edu/abs/2005MNRAS.358..551B}{358}}, 551

\bibitem[{{Bett} \& {Frenk}(2012)}]{Bett2012}
{Bett}, P.~E., \& {Frenk}, C.~S. 2012,
  {\href{https://doi.org/10.1111/j.1365-2966.2011.20275.x}{\mnras}},
  {\href{https://ui.adsabs.harvard.edu/abs/2012MNRAS.420.3324B}{420}}, 3324

\bibitem[{{Blake} {et~al.}(2011){Blake}, {Brough}, {Colless}, {Contreras},
  {Couch}, {Croom}, {Davis}, {Drinkwater}, {Forster}, {Gilbank}, {Gladders},
  {Glazebrook}, {Jelliffe}, {Jurek}, {Li}, {Madore}, {Martin}, {Pimbblet},
  {Poole}, {Pracy}, {Sharp}, {Wisnioski}, {Woods}, {Wyder}, \&
  {Yee}}]{Blake2011}
{Blake}, C., {Brough}, S., {Colless}, M., {et~al.} 2011,
  {\href{https://doi.org/10.1111/j.1365-2966.2011.18903.x}{\mnras}},
  {\href{https://ui.adsabs.harvard.edu/abs/2011MNRAS.415.2876B}{415}}, 2876

\bibitem[{{Bond} {et~al.}(1991){Bond}, {Cole}, {Efstathiou}, \&
  {Kaiser}}]{Bond1991}
{Bond}, J.~R., {Cole}, S., {Efstathiou}, G., \& {Kaiser}, N. 1991,
  {\href{https://doi.org/10.1086/170520}{\apj}},
  {\href{https://ui.adsabs.harvard.edu/abs/1991ApJ...379..440B}{379}}, 440

\bibitem[{{Bryan} \& {Norman}(1998)}]{Bryan1998}
{Bryan}, G.~L., \& {Norman}, M.~L. 1998,
  {\href{https://doi.org/10.1086/305262}{\apj}},
  {\href{https://ui.adsabs.harvard.edu/abs/1998ApJ...495...80B}{495}}, 80

\bibitem[{{Bullock} {et~al.}(2001){Bullock}, {Dekel}, {Kolatt}, {Kravtsov},
  {Klypin}, {Porciani}, \& {Primack}}]{Bullock2001b}
{Bullock}, J.~S., {Dekel}, A., {Kolatt}, T.~S., {et~al.} 2001,
  {\href{https://doi.org/10.1086/321477}{\apj}},
  {\href{https://ui.adsabs.harvard.edu/abs/2001ApJ...555..240B}{555}}, 240

\bibitem[{{Chen} {et~al.}(2016){Chen}, {Wang}, {Mo}, \& {Shi}}]{Chen2016}
{Chen}, S., {Wang}, H., {Mo}, H.~J., \& {Shi}, J. 2016,
  {\href{https://doi.org/10.3847/0004-637X/825/1/49}{\apj}},
  {\href{https://ui.adsabs.harvard.edu/abs/2016ApJ...825...49C}{825}}, 49

\bibitem[{{Dalal} {et~al.}(2010){Dalal}, {Lithwick}, \& {Kuhlen}}]{Dalal2010}
{Dalal}, N., {Lithwick}, Y., \& {Kuhlen}, M. 2010,
  {arXiv:\href{https://arxiv.org/abs/1010.2539}{1010.2539}}

\bibitem[{{Davis} {et~al.}(1985){Davis}, {Efstathiou}, {Frenk}, \&
  {White}}]{Davis1985}
{Davis}, M., {Efstathiou}, G., {Frenk}, C.~S., \& {White}, S.~D.~M. 1985,
  {\href{https://doi.org/10.1086/163168}{\apj}},
  {\href{https://ui.adsabs.harvard.edu/abs/1985ApJ...292..371D}{292}}, 371

\bibitem[{{Despali} {et~al.}(2016){Despali}, {Giocoli}, {Angulo}, {Tormen},
  {Sheth}, {Baso}, \& {Moscardini}}]{Despali2016}
{Despali}, G., {Giocoli}, C., {Angulo}, R.~E., {et~al.} 2016,
  {\href{https://doi.org/10.1093/mnras/stv2842}{\mnras}},
  {\href{https://ui.adsabs.harvard.edu/abs/2016MNRAS.456.2486D}{456}}, 2486

\bibitem[{{Diaferio} \& {Geller}(1997)}]{Diaferio1997}
{Diaferio}, A., \& {Geller}, M.~J. 1997,
  {\href{https://doi.org/10.1086/304075}{\apj}},
  {\href{https://ui.adsabs.harvard.edu/abs/1997ApJ...481..633D}{481}}, 633

\bibitem[{{Diemer}(2018)}]{Diemer2017a}
{Diemer}, B. 2018, {\href{https://doi.org/10.3847/1538-4365/aaee8c}{\apjs}},
  {\href{https://ui.adsabs.harvard.edu/abs/2018ApJS..239...35D}{239}}, 35

\bibitem[{{Diemer}(2020)}]{Diemer2020}
---. 2020, {\href{https://doi.org/10.3847/1538-4357/abbf52}{\apj}},
  {\href{https://ui.adsabs.harvard.edu/abs/2020ApJ...903...87D}{903}}, 87

\bibitem[{{Fakhouri} {et~al.}(2010){Fakhouri}, {Ma}, \&
  {Boylan-Kolchin}}]{Fakhouri2010}
{Fakhouri}, O., {Ma}, C.-P., \& {Boylan-Kolchin}, M. 2010,
  {\href{https://doi.org/10.1111/j.1365-2966.2010.16859.x}{\mnras}},
  {\href{https://ui.adsabs.harvard.edu/abs/2010MNRAS.406.2267F}{406}}, 2267

\bibitem[{{Fillingham} {et~al.}(2015){Fillingham}, {Cooper}, {Wheeler},
  {Garrison-Kimmel}, {Boylan-Kolchin}, \& {Bullock}}]{Fillingham2015}
{Fillingham}, S.~P., {Cooper}, M.~C., {Wheeler}, C., {et~al.} 2015,
  {\href{https://doi.org/10.1093/mnras/stv2058}{\mnras}},
  {\href{https://ui.adsabs.harvard.edu/abs/2015MNRAS.454.2039F}{454}}, 2039

\bibitem[{{Fujita}(2004)}]{Fujita2004}
{Fujita}, Y. 2004, {\href{https://doi.org/10.1093/pasj/56.1.29}{\pasj}},
  {\href{https://ui.adsabs.harvard.edu/abs/2004PASJ...56...29F}{56}}, 29

\bibitem[{{Gao} {et~al.}(2004){Gao}, {White}, {Jenkins}, {Stoehr}, \&
  {Springel}}]{Gao2004}
{Gao}, L., {White}, S.~D.~M., {Jenkins}, A., {Stoehr}, F., \& {Springel}, V.
  2004, {\href{https://doi.org/10.1111/j.1365-2966.2004.08360.x}{\mnras}},
  {\href{https://ui.adsabs.harvard.edu/abs/2004MNRAS.355..819G}{355}}, 819

\bibitem[{{Gill} {et~al.}(2005){Gill}, {Knebe}, \& {Gibson}}]{Gill2005}
{Gill}, S. P.~D., {Knebe}, A., \& {Gibson}, B.~K. 2005,
  {\href{https://doi.org/10.1111/j.1365-2966.2004.08562.x}{\mnras}},
  {\href{https://ui.adsabs.harvard.edu/abs/2005MNRAS.356.1327G}{356}}, 1327

\bibitem[{{Gunn} \& {Gott}(1972)}]{Gunn1972}
{Gunn}, J.~E., \& {Gott}, J.~Richard, I. 1972,
  {\href{https://doi.org/10.1086/151605}{\apj}},
  {\href{https://ui.adsabs.harvard.edu/abs/1972ApJ...176....1G}{176}}, 1

\bibitem[{{Han} {et~al.}(2018){Han}, {Cole}, {Frenk}, {Benitez-Llambay}, \&
  {Helly}}]{Han2018a}
{Han}, J., {Cole}, S., {Frenk}, C.~S., {Benitez-Llambay}, A., \& {Helly}, J.
  2018, {\href{https://doi.org/10.1093/mnras/stx2792}{\mnras}},
  {\href{https://ui.adsabs.harvard.edu/abs/2018MNRAS.474..604H}{474}}, 604

\bibitem[{{Han} {et~al.}(2016){Han}, {Cole}, {Frenk}, \& {Jing}}]{Han2016}
{Han}, J., {Cole}, S., {Frenk}, C.~S., \& {Jing}, Y. 2016,
  {\href{https://doi.org/10.1093/mnras/stv2900}{\mnras}},
  {\href{https://ui.adsabs.harvard.edu/abs/2016MNRAS.457.1208H}{457}}, 1208

\bibitem[{{Han} {et~al.}(2012){Han}, {Jing}, {Wang}, \& {Wang}}]{Han2012}
{Han}, J., {Jing}, Y.~P., {Wang}, H., \& {Wang}, W. 2012,
  {\href{https://doi.org/10.1111/j.1365-2966.2012.22111.x}{\mnras}},
  {\href{https://ui.adsabs.harvard.edu/abs/2012MNRAS.427.2437H}{427}}, 2437

\bibitem[{{Hunter}(2007)}]{Hunter2007}
{Hunter}, J.~D. 2007, {\href{https://doi.org/10.1109/MCSE.2007.55}{CSE}},
  {\href{https://ui.adsabs.harvard.edu/abs/2007CSE.....9...90H}{9}}, 90

\bibitem[{{Ibata} {et~al.}(2013){Ibata}, {Lewis}, {Conn}, {Irwin},
  {McConnachie}, {Chapman}, {Collins}, {Fardal}, {Ferguson}, {Ibata}, {Mackey},
  {Martin}, {Navarro}, {Rich}, {Valls-Gabaud}, \& {Widrow}}]{Ibata2013}
{Ibata}, R.~A., {Lewis}, G.~F., {Conn}, A.~R., {et~al.} 2013,
  {\href{https://doi.org/10.1038/nature11717}{\nat}},
  {\href{https://ui.adsabs.harvard.edu/abs/2013Natur.493...62I}{493}}, 62

\bibitem[{{Jiang} {et~al.}(2008){Jiang}, {Jing}, {Faltenbacher}, {Lin}, \&
  {Li}}]{Jiang2008}
{Jiang}, C.~Y., {Jing}, Y.~P., {Faltenbacher}, A., {Lin}, W.~P., \& {Li}, C.
  2008, {\href{https://doi.org/10.1086/526412}{\apj}},
  {\href{https://ui.adsabs.harvard.edu/abs/2008ApJ...675.1095J}{675}}, 1095

\bibitem[{{Jiang} {et~al.}(2020){Jiang}, {Dekel}, {Freundlich}, {van den
  Bosch}, {Green}, {Hopkins}, {Benson}, \& {Du}}]{Jiang2020}
{Jiang}, F., {Dekel}, A., {Freundlich}, J., {et~al.} 2020,
  {arXiv:\href{https://arxiv.org/abs/2005.05974}{2005.05974}}

\bibitem[{{Jiang} \& {van den Bosch}(2017)}]{Jiang2016}
{Jiang}, F., \& {van den Bosch}, F.~C. 2017,
  {\href{https://doi.org/10.1093/mnras/stx1979}{\mnras}},
  {\href{https://ui.adsabs.harvard.edu/abs/2017MNRAS.472..657J}{472}}, 657

\bibitem[{{Jiang} {et~al.}(2015){Jiang}, {Cole}, {Sawala}, \&
  {Frenk}}]{Jiang2015}
{Jiang}, L., {Cole}, S., {Sawala}, T., \& {Frenk}, C.~S. 2015,
  {\href{https://doi.org/10.1093/mnras/stv053}{\mnras}},
  {\href{https://ui.adsabs.harvard.edu/abs/2015MNRAS.448.1674J}{448}}, 1674

\bibitem[{{Jing} \& {Suto}(2002)}]{Jing2002}
{Jing}, Y.~P., \& {Suto}, Y. 2002,
  {\href{https://doi.org/10.1086/341065}{\apj}},
  {\href{https://ui.adsabs.harvard.edu/abs/2002ApJ...574..538J}{574}}, 538

\bibitem[{{Jing} {et~al.}(2007){Jing}, {Suto}, \& {Mo}}]{Jing2007}
{Jing}, Y.~P., {Suto}, Y., \& {Mo}, H.~J. 2007,
  {\href{https://doi.org/10.1086/511130}{\apj}},
  {\href{https://ui.adsabs.harvard.edu/abs/2007ApJ...657..664J}{657}}, 664

\bibitem[{{Kang} \& {Wang}(2015)}]{Kang2015a}
{Kang}, X., \& {Wang}, P. 2015,
  {\href{https://doi.org/10.1088/0004-637X/813/1/6}{\apj}},
  {\href{https://ui.adsabs.harvard.edu/abs/2015ApJ...813....6K}{813}}, 6

\bibitem[{{Kroupa} {et~al.}(2005){Kroupa}, {Theis}, \& {Boily}}]{Kroupa2005}
{Kroupa}, P., {Theis}, C., \& {Boily}, C.~M. 2005,
  {\href{https://doi.org/10.1051/0004-6361:20041122}{\aap}},
  {\href{https://ui.adsabs.harvard.edu/abs/2005A&A...431..517K}{431}}, 517

\bibitem[{{Lacey} \& {Cole}(1993)}]{Lacey1993}
{Lacey}, C., \& {Cole}, S. 1993,
  {\href{https://doi.org/10.1093/mnras/262.3.627}{\mnras}},
  {\href{https://ui.adsabs.harvard.edu/abs/1993MNRAS.262..627L}{262}}, 627

\bibitem[{{Lemze} {et~al.}(2012){Lemze}, {Wagner}, {Rephaeli}, {Sadeh},
  {Norman}, {Barkana}, {Broadhurst}, {Ford}, \& {Postman}}]{Lemze2012}
{Lemze}, D., {Wagner}, R., {Rephaeli}, Y., {et~al.} 2012,
  {\href{https://doi.org/10.1088/0004-637X/752/2/141}{\apj}},
  {\href{https://ui.adsabs.harvard.edu/abs/2012ApJ...752..141L}{752}}, 141

\bibitem[{{Li} {et~al.}(2017){Li}, {Jing}, {Qian}, {Yuan}, \& {Zhao}}]{Li2017}
{Li}, Z.-Z., {Jing}, Y.~P., {Qian}, Y.-Z., {Yuan}, Z., \& {Zhao}, D.-H. 2017,
  {\href{https://doi.org/10.3847/1538-4357/aa94c0}{\apj}},
  {\href{https://ui.adsabs.harvard.edu/abs/2017ApJ...850..116L}{850}}, 116

\bibitem[{{Li} {et~al.}(2019){Li}, {Qian}, {Han}, {Wang}, \& {Jing}}]{Li2019a}
{Li}, Z.-Z., {Qian}, Y.-Z., {Han}, J., {Wang}, W., \& {Jing}, Y.~P. 2019,
  {\href{https://doi.org/10.3847/1538-4357/ab4f6d}{\apj}},
  {\href{https://ui.adsabs.harvard.edu/abs/2019ApJ...886...69L}{886}}, 69

\bibitem[{{Libeskind} {et~al.}(2014){Libeskind}, {Knebe}, {Hoffman}, \&
  {Gottl{\"o}ber}}]{Libeskind2014}
{Libeskind}, N.~I., {Knebe}, A., {Hoffman}, Y., \& {Gottl{\"o}ber}, S. 2014,
  {\href{https://doi.org/10.1093/mnras/stu1216}{\mnras}},
  {\href{https://ui.adsabs.harvard.edu/abs/2014MNRAS.443.1274L}{443}}, 1274

\bibitem[{{Ludlow} {et~al.}(2009){Ludlow}, {Navarro}, {Springel}, {Jenkins},
  {Frenk}, \& {Helmi}}]{Ludlow2009}
{Ludlow}, A.~D., {Navarro}, J.~F., {Springel}, V., {et~al.} 2009,
  {\href{https://doi.org/10.1088/0004-637X/692/1/931}{\apj}},
  {\href{https://ui.adsabs.harvard.edu/abs/2009ApJ...692..931L}{692}}, 931

\bibitem[{{Ludlow} {et~al.}(2011){Ludlow}, {Navarro}, {White},
  {Boylan-Kolchin}, {Springel}, {Jenkins}, \& {Frenk}}]{Ludlow2011}
{Ludlow}, A.~D., {Navarro}, J.~F., {White}, S. D.~M., {et~al.} 2011,
  {\href{https://doi.org/10.1111/j.1365-2966.2011.19008.x}{\mnras}},
  {\href{https://ui.adsabs.harvard.edu/abs/2011MNRAS.415.3895L}{415}}, 3895

\bibitem[{{Mamon} {et~al.}(2004){Mamon}, {Sanchis}, {Salvador-Sol{\'e}}, \&
  {Solanes}}]{Mamon2004}
{Mamon}, G.~A., {Sanchis}, T., {Salvador-Sol{\'e}}, E., \& {Solanes}, J.~M.
  2004, {\href{https://doi.org/10.1051/0004-6361:20034155}{\aap}},
  {\href{https://ui.adsabs.harvard.edu/abs/2004A&A...414..445M}{414}}, 445

\bibitem[{{Martin} {et~al.}(2020){Martin}, {Jackson}, {Kaviraj}, {Choi},
  {Devriendt}, {Dubois}, {Kimm}, {Kraljic}, {Peirani}, {Pichon}, {Volonteri},
  \& {Yi}}]{Martin2020}
{Martin}, G., {Jackson}, R.~A., {Kaviraj}, S., {et~al.} 2020,
  {\href{https://doi.org/10.1093/mnras/staa3443}{\mnras}} in press,
  arXiv:2007.07913

\bibitem[{{McGee} {et~al.}(2009){McGee}, {Balogh}, {Bower}, {Font}, \&
  {McCarthy}}]{McGee2009}
{McGee}, S.~L., {Balogh}, M.~L., {Bower}, R.~G., {Font}, A.~S., \& {McCarthy},
  I.~G. 2009,
  {\href{https://doi.org/10.1111/j.1365-2966.2009.15507.x}{\mnras}},
  {\href{https://ui.adsabs.harvard.edu/abs/2009MNRAS.400..937M}{400}}, 937

\bibitem[{Mo {et~al.}(2010)Mo, van~den Bosch, \& White}]{Mo2010}
Mo, H., van~den Bosch, F., \& White, S. 2010, Galaxy {{Formation}} and
  {{Evolution}} ({Cambridge University Press})

\bibitem[{{More} {et~al.}(2011){More}, {Kravtsov}, {Dalal}, \&
  {Gottl{\"o}ber}}]{More2011}
{More}, S., {Kravtsov}, A.~V., {Dalal}, N., \& {Gottl{\"o}ber}, S. 2011,
  {\href{https://doi.org/10.1088/0067-0049/195/1/4}{\apjs}},
  {\href{https://ui.adsabs.harvard.edu/abs/2011ApJS..195....4M}{195}}, 4

\bibitem[{{Morinaga} \& {Ishiyama}(2020)}]{Morinaga2020}
{Morinaga}, Y., \& {Ishiyama}, T. 2020,
  {\href{https://doi.org/10.1093/mnras/staa1180}{\mnras}},
  {\href{https://ui.adsabs.harvard.edu/abs/2020MNRAS.495..502M}{495}}, 502

\bibitem[{{Navarro} {et~al.}(1996){Navarro}, {Frenk}, \& {White}}]{Navarro1996}
{Navarro}, J.~F., {Frenk}, C.~S., \& {White}, S. D.~M. 1996,
  {\href{https://doi.org/10.1086/177173}{\apj}},
  {\href{https://ui.adsabs.harvard.edu/abs/1996ApJ...462..563N}{462}}, 563

\bibitem[{{Navarro} {et~al.}(2004){Navarro}, {Hayashi}, {Power}, {Jenkins},
  {Frenk}, {White}, {Springel}, {Stadel}, \& {Quinn}}]{Navarro2004}
{Navarro}, J.~F., {Hayashi}, E., {Power}, C., {et~al.} 2004,
  {\href{https://doi.org/10.1111/j.1365-2966.2004.07586.x}{\mnras}},
  {\href{https://ui.adsabs.harvard.edu/abs/2004MNRAS.349.1039N}{349}}, 1039

\bibitem[{{Navarro} {et~al.}(2010){Navarro}, {Ludlow}, {Springel}, {Wang},
  {Vogelsberger}, {White}, {Jenkins}, {Frenk}, \& {Helmi}}]{Navarro2010}
{Navarro}, J.~F., {Ludlow}, A., {Springel}, V., {et~al.} 2010,
  {\href{https://doi.org/10.1111/j.1365-2966.2009.15878.x}{\mnras}},
  {\href{https://ui.adsabs.harvard.edu/abs/2010MNRAS.402...21N}{402}}, 21

\bibitem[{{Oliphant}(2007)}]{Oliphant2007}
{Oliphant}, T.~E. 2007, {\href{https://doi.org/10.1109/MCSE.2007.58}{CSE}},
  {\href{https://ui.adsabs.harvard.edu/abs/2007CSE.....9c..10O}{9}}, 10

\bibitem[{{Onions} {et~al.}(2012){Onions}, {Knebe}, {Pearce}, {Muldrew}, {Lux},
  {Knollmann}, {Ascasibar}, {Behroozi}, {Elahi}, {Han}, {Maciejewski},
  {Merch{\'a}n}, {Neyrinck}, {Ruiz}, {Sgr{\'o}}, {Springel}, \&
  {Tweed}}]{Onions2012}
{Onions}, J., {Knebe}, A., {Pearce}, F.~R., {et~al.} 2012,
  {\href{https://doi.org/10.1111/j.1365-2966.2012.20947.x}{\mnras}},
  {\href{https://ui.adsabs.harvard.edu/abs/2012MNRAS.423.1200O}{423}}, 1200

\bibitem[{{Pawlowski} {et~al.}(2012){Pawlowski}, {Pflamm-Altenburg}, \&
  {Kroupa}}]{Pawlowski2012}
{Pawlowski}, M.~S., {Pflamm-Altenburg}, J., \& {Kroupa}, P. 2012,
  {\href{https://doi.org/10.1111/j.1365-2966.2012.20937.x}{\mnras}},
  {\href{https://ui.adsabs.harvard.edu/abs/2012MNRAS.423.1109P}{423}}, 1109

\bibitem[{{Press} \& {Schechter}(1974)}]{Press1974}
{Press}, W.~H., \& {Schechter}, P. 1974,
  {\href{https://doi.org/10.1086/152650}{\apj}},
  {\href{https://ui.adsabs.harvard.edu/abs/1974ApJ...187..425P}{187}}, 425

\bibitem[{{Richings} {et~al.}(2020){Richings}, {Frenk}, {Jenkins}, {Robertson},
  {Fattahi}, {Grand}, {Navarro}, {Pakmor}, {Gomez}, {Marinacci}, \&
  {Oman}}]{Richings2018}
{Richings}, J., {Frenk}, C., {Jenkins}, A., {et~al.} 2020,
  {\href{https://doi.org/10.1093/mnras/stz3448}{\mnras}},
  {\href{https://ui.adsabs.harvard.edu/abs/2020MNRAS.492.5780R}{492}}, 5780

\bibitem[{{Sales} {et~al.}(2007){Sales}, {Navarro}, {Abadi}, \&
  {Steinmetz}}]{Sales2007}
{Sales}, L.~V., {Navarro}, J.~F., {Abadi}, M.~G., \& {Steinmetz}, M. 2007,
  {\href{https://doi.org/10.1111/j.1365-2966.2007.12026.x}{\mnras}},
  {\href{https://ui.adsabs.harvard.edu/abs/2007MNRAS.379.1475S}{379}}, 1475

\bibitem[{{Sawala} {et~al.}(2017){Sawala}, {Pihajoki}, {Johansson}, {Frenk},
  {Navarro}, {Oman}, \& {White}}]{Sawala2017}
{Sawala}, T., {Pihajoki}, P., {Johansson}, P.~H., {et~al.} 2017,
  {\href{https://doi.org/10.1093/mnras/stx360}{\mnras}},
  {\href{https://ui.adsabs.harvard.edu/abs/2017MNRAS.467.4383S}{467}}, 4383

\bibitem[{{Shao} {et~al.}(2018){Shao}, {Cautun}, {Frenk}, {Grand}, {G{\'o}mez},
  {Marinacci}, \& {Simpson}}]{Shao2018a}
{Shao}, S., {Cautun}, M., {Frenk}, C.~S., {et~al.} 2018,
  {\href{https://doi.org/10.1093/mnras/sty343}{\mnras}},
  {\href{https://ui.adsabs.harvard.edu/abs/2018MNRAS.476.1796S}{476}}, 1796

\bibitem[{{Sheth} {et~al.}(2001){Sheth}, {Mo}, \& {Tormen}}]{Sheth2001}
{Sheth}, R.~K., {Mo}, H.~J., \& {Tormen}, G. 2001,
  {\href{https://doi.org/10.1046/j.1365-8711.2001.04006.x}{\mnras}},
  {\href{https://ui.adsabs.harvard.edu/abs/2001MNRAS.323....1S}{323}}, 1

\bibitem[{{Shi} {et~al.}(2016){Shi}, {Yang}, {Wang}, {Zhang}, {Mo}, {van den
  Bosch}, {Li}, {Liu}, {Lu}, {Tweed}, \& {Yang}}]{Shi2016}
{Shi}, F., {Yang}, X., {Wang}, H., {et~al.} 2016,
  {\href{https://doi.org/10.3847/1538-4357/833/2/241}{\apj}},
  {\href{https://ui.adsabs.harvard.edu/abs/2016ApJ...833..241S}{833}}, 241

\bibitem[{{Shi} {et~al.}(2015){Shi}, {Wang}, \& {Mo}}]{Shi2015}
{Shi}, J., {Wang}, H., \& {Mo}, H.~J. 2015,
  {\href{https://doi.org/10.1088/0004-637X/807/1/37}{\apj}},
  {\href{https://ui.adsabs.harvard.edu/abs/2015ApJ...807...37S}{807}}, 37

\bibitem[{{Sparre} \& {Hansen}(2012)}]{Sparre2012}
{Sparre}, M., \& {Hansen}, S.~H. 2012,
  {\href{https://doi.org/10.1088/1475-7516/2012/07/042}{\jcap}},
  {\href{https://ui.adsabs.harvard.edu/abs/2012JCAP...07..042S}{2012}}, 042

\bibitem[{{Springel} {et~al.}(2001){Springel}, {White}, {Tormen}, \&
  {Kauffmann}}]{Springel2001}
{Springel}, V., {White}, S. D.~M., {Tormen}, G., \& {Kauffmann}, G. 2001,
  {\href{https://doi.org/10.1046/j.1365-8711.2001.04912.x}{\mnras}},
  {\href{https://ui.adsabs.harvard.edu/abs/2001MNRAS.328..726S}{328}}, 726

\bibitem[{{Springel} {et~al.}(2008){Springel}, {Wang}, {Vogelsberger},
  {Ludlow}, {Jenkins}, {Helmi}, {Navarro}, {Frenk}, \& {White}}]{Springel2008}
{Springel}, V., {Wang}, J., {Vogelsberger}, M., {et~al.} 2008,
  {\href{https://doi.org/10.1111/j.1365-2966.2008.14066.x}{\mnras}},
  {\href{https://ui.adsabs.harvard.edu/abs/2008MNRAS.391.1685S}{391}}, 1685

\bibitem[{{Srisawat} {et~al.}(2013){Srisawat}, {Knebe}, {Pearce}, {Schneider},
  {Thomas}, {Behroozi}, {Dolag}, {Elahi}, {Han}, {Helly}, {Jing}, {Jung},
  {Lee}, {Mao}, {Onions}, {Rodriguez-Gomez}, {Tweed}, \& {Yi}}]{Srisawat2013}
{Srisawat}, C., {Knebe}, A., {Pearce}, F.~R., {et~al.} 2013,
  {\href{https://doi.org/10.1093/mnras/stt1545}{\mnras}},
  {\href{https://ui.adsabs.harvard.edu/abs/2013MNRAS.436..150S}{436}}, 150

\bibitem[{{Taylor} \& {Navarro}(2001)}]{Taylor2001}
{Taylor}, J.~E., \& {Navarro}, J.~F. 2001,
  {\href{https://doi.org/10.1086/324031}{\apj}},
  {\href{https://ui.adsabs.harvard.edu/abs/2001ApJ...563..483T}{563}}, 483

\bibitem[{Tormen(1997)}]{Tormen1997}
Tormen, G. 1997, MNRAS, 290, 411

\bibitem[{{van den Bosch}(2017)}]{vandenBosch2017}
{van den Bosch}, F.~C. 2017,
  {\href{https://doi.org/10.1093/mnras/stx520}{\mnras}},
  {\href{https://ui.adsabs.harvard.edu/abs/2017MNRAS.468..885V}{468}}, 885

\bibitem[{{van den Bosch} {et~al.}(2014){van den Bosch}, {Jiang}, {Hearin},
  {Campbell}, {Watson}, \& {Padmanabhan}}]{vandenBosch2014}
{van den Bosch}, F.~C., {Jiang}, F., {Hearin}, A., {et~al.} 2014,
  {\href{https://doi.org/10.1093/mnras/stu1872}{\mnras}},
  {\href{https://ui.adsabs.harvard.edu/abs/2014MNRAS.445.1713V}{445}}, 1713

\bibitem[{{van den Bosch} \& {Ogiya}(2018)}]{vandenBosch2018a}
{van den Bosch}, F.~C., \& {Ogiya}, G. 2018,
  {\href{https://doi.org/10.1093/mnras/sty084}{\mnras}},
  {\href{https://ui.adsabs.harvard.edu/abs/2018MNRAS.475.4066V}{475}}, 4066

\bibitem[{{van der Walt} {et~al.}(2011){van der Walt}, {Colbert}, \&
  {Varoquaux}}]{Walt2011}
{van der Walt}, S., {Colbert}, S.~C., \& {Varoquaux}, G. 2011,
  {\href{https://doi.org/10.1109/MCSE.2011.37}{CSE}},
  {\href{https://ui.adsabs.harvard.edu/abs/2011CSE....13b..22V}{13}}, 22

\bibitem[{{Vijayaraghavan} \& {Ricker}(2013)}]{Vijayaraghavan2013}
{Vijayaraghavan}, R., \& {Ricker}, P.~M. 2013,
  {\href{https://doi.org/10.1093/mnras/stt1485}{\mnras}},
  {\href{https://ui.adsabs.harvard.edu/abs/2013MNRAS.435.2713V}{435}}, 2713

\bibitem[{{Vitvitska} {et~al.}(2002){Vitvitska}, {Klypin}, {Kravtsov},
  {Wechsler}, {Primack}, \& {Bullock}}]{Vitvitska2002}
{Vitvitska}, M., {Klypin}, A.~A., {Kravtsov}, A.~V., {et~al.} 2002,
  {\href{https://doi.org/10.1086/344361}{\apj}},
  {\href{https://ui.adsabs.harvard.edu/abs/2002ApJ...581..799V}{581}}, 799

\bibitem[{{Wang} {et~al.}(2009){Wang}, {Mo}, \& {Jing}}]{Wang2009}
{Wang}, H., {Mo}, H.~J., \& {Jing}, Y.~P. 2009,
  {\href{https://doi.org/10.1111/j.1365-2966.2009.14884.x}{\mnras}},
  {\href{https://ui.adsabs.harvard.edu/abs/2009MNRAS.396.2249W}{396}}, 2249

\bibitem[{{Wang} {et~al.}(2011){Wang}, {Mo}, {Jing}, {Yang}, \&
  {Wang}}]{Wang2011}
{Wang}, H., {Mo}, H.~J., {Jing}, Y.~P., {Yang}, X., \& {Wang}, Y. 2011,
  {\href{https://doi.org/10.1111/j.1365-2966.2011.18301.x}{\mnras}},
  {\href{https://ui.adsabs.harvard.edu/abs/2011MNRAS.413.1973W}{413}}, 1973

\bibitem[{{Wang} {et~al.}(2005){Wang}, {Jing}, {Mao}, \& {Kang}}]{Wang2005}
{Wang}, H.~Y., {Jing}, Y.~P., {Mao}, S., \& {Kang}, X. 2005,
  {\href{https://doi.org/10.1111/j.1365-2966.2005.09543.x}{\mnras}},
  {\href{https://ui.adsabs.harvard.edu/abs/2005MNRAS.364..424W}{364}}, 424

\bibitem[{{Wang} \& {Kang}(2018)}]{Wang2017a}
{Wang}, P., \& {Kang}, X. 2018,
  {\href{https://doi.org/10.1093/mnras/stx2466}{\mnras}},
  {\href{https://ui.adsabs.harvard.edu/abs/2018MNRAS.473.1562W}{473}}, 1562

\bibitem[{{Wang} {et~al.}(2020){Wang}, {Libeskind}, {Tempel}, {Pawlowski},
  {Kang}, \& {Guo}}]{Wang2020a}
{Wang}, P., {Libeskind}, N.~I., {Tempel}, E., {et~al.} 2020,
  {\href{https://doi.org/10.3847/1538-4357/aba6ea}{\apj}},
  {\href{https://ui.adsabs.harvard.edu/abs/2020ApJ...900..129W}{900}}, 129

\bibitem[{{Wetzel}(2011)}]{Wetzel2011}
{Wetzel}, A.~R. 2011,
  {\href{https://doi.org/10.1111/j.1365-2966.2010.17877.x}{\mnras}},
  {\href{https://ui.adsabs.harvard.edu/abs/2011MNRAS.412...49W}{412}}, 49

\bibitem[{{Wetzel} {et~al.}(2013){Wetzel}, {Tinker}, {Conroy}, \& {van den
  Bosch}}]{Wetzel2013}
{Wetzel}, A.~R., {Tinker}, J.~L., {Conroy}, C., \& {van den Bosch}, F.~C. 2013,
  {\href{https://doi.org/10.1093/mnras/stt469}{\mnras}},
  {\href{https://ui.adsabs.harvard.edu/abs/2013MNRAS.432..336W}{432}}, 336

\bibitem[{{White}(1984)}]{White1984}
{White}, S.~D.~M. 1984, {\href{https://doi.org/10.1086/162573}{\apj}},
  {\href{https://ui.adsabs.harvard.edu/abs/1984ApJ...286...38W}{286}}, 38

\bibitem[{{Yang} {et~al.}(2011){Yang}, {Mo}, {Zhang}, \& {van den
  Bosch}}]{Yang2011}
{Yang}, X., {Mo}, H.~J., {Zhang}, Y., \& {van den Bosch}, F.~C. 2011,
  {\href{https://doi.org/10.1088/0004-637X/741/1/13}{\apj}},
  {\href{https://ui.adsabs.harvard.edu/abs/2011ApJ...741...13Y}{741}}, 13

\bibitem[{{Yang} {et~al.}(2006){Yang}, {van den Bosch}, {Mo}, {Mao}, {Kang},
  {Weinmann}, {Guo}, \& {Jing}}]{Yang2006}
{Yang}, X., {van den Bosch}, F.~C., {Mo}, H.~J., {et~al.} 2006,
  {\href{https://doi.org/10.1111/j.1365-2966.2006.10373.x}{\mnras}},
  {\href{https://ui.adsabs.harvard.edu/abs/2006MNRAS.369.1293Y}{369}}, 1293

\bibitem[{{Zavala} \& {Frenk}(2019)}]{Zavala2019}
{Zavala}, J., \& {Frenk}, C.~S. 2019,
  {\href{https://doi.org/10.3390/galaxies7040081}{Galax}},
  {\href{https://ui.adsabs.harvard.edu/abs/2019Galax...7...81Z}{7}}, 81

\bibitem[{Zel'dovich(1970)}]{Zeldovich1970}
Zel'dovich, Y.~B. 1970, A\&A, 5, 84

\bibitem[{{Zhao} {et~al.}(2009){Zhao}, {Jing}, {Mo}, \&
  {B{\"o}rner}}]{Zhao2009}
{Zhao}, D.~H., {Jing}, Y.~P., {Mo}, H.~J., \& {B{\"o}rner}, G. 2009,
  {\href{https://doi.org/10.1088/0004-637X/707/1/354}{\apj}},
  {\href{https://ui.adsabs.harvard.edu/abs/2009ApJ...707..354Z}{707}}, 354

\bibitem[{{Zhu} {et~al.}(2016){Zhu}, {Marinacci}, {Maji}, {Li}, {Springel}, \&
  {Hernquist}}]{Zhu2016a}
{Zhu}, Q., {Marinacci}, F., {Maji}, M., {et~al.} 2016,
  {\href{https://doi.org/10.1093/mnras/stw374}{\mnras}},
  {\href{https://ui.adsabs.harvard.edu/abs/2016MNRAS.458.1559Z}{458}}, 1559

\bibitem[{{Zu} \& {Weinberg}(2013)}]{Zu2013}
{Zu}, Y., \& {Weinberg}, D.~H. 2013,
  {\href{https://doi.org/10.1093/mnras/stt411}{\mnras}},
  {\href{https://ui.adsabs.harvard.edu/abs/2013MNRAS.431.3319Z}{431}}, 3319

\end{thebibliography}

\end{document}